\PassOptionsToPackage{unicode}{hyperref}
\PassOptionsToPackage{hyphens}{url}
\PassOptionsToPackage{dvipsnames,svgnames,x11names}{xcolor}
\documentclass[
  12pt]{article}

\usepackage{amsmath,amssymb}
\usepackage{iftex}
\ifPDFTeX
  \usepackage[T1]{fontenc}
  \usepackage[utf8]{inputenc}
  \usepackage{textcomp} % provide euro and other symbols
\else % if luatex or xetex
  \usepackage{unicode-math}
  \defaultfontfeatures{Scale=MatchLowercase}
  \defaultfontfeatures[\rmfamily]{Ligatures=TeX,Scale=1}
\fi
\usepackage{lmodern}
\ifPDFTeX\else  
\fi
\IfFileExists{upquote.sty}{\usepackage{upquote}}{}
\IfFileExists{microtype.sty}{% use microtype if available
  \usepackage[]{microtype}
  \UseMicrotypeSet[protrusion]{basicmath} % disable protrusion for tt fonts
}{}
\makeatletter
\@ifundefined{KOMAClassName}{% if non-KOMA class
  \IfFileExists{parskip.sty}{%
    \usepackage{parskip}
  }{% else
    \setlength{\parindent}{0pt}
    \setlength{\parskip}{6pt plus 2pt minus 1pt}}
}{% if KOMA class
  \KOMAoptions{parskip=half}}
\makeatother
\usepackage{xcolor}
\makeatletter
\ifx\paragraph\undefined\else
  \let\oldparagraph\paragraph
  \renewcommand{\paragraph}{
    \@ifstar
      \xxxParagraphStar
      \xxxParagraphNoStar
  }
  \newcommand{\xxxParagraphStar}[1]{\oldparagraph*{#1}\mbox{}}
  \newcommand{\xxxParagraphNoStar}[1]{\oldparagraph{#1}\mbox{}}
\fi
\ifx\subparagraph\undefined\else
  \let\oldsubparagraph\subparagraph
  \renewcommand{\subparagraph}{
    \@ifstar
      \xxxSubParagraphStar
      \xxxSubParagraphNoStar
  }
  \newcommand{\xxxSubParagraphStar}[1]{\oldsubparagraph*{#1}\mbox{}}
  \newcommand{\xxxSubParagraphNoStar}[1]{\oldsubparagraph{#1}\mbox{}}
\fi
\makeatother

\usepackage{longtable,booktabs,array}
\usepackage{calc} % for calculating minipage widths
\usepackage{etoolbox}
\makeatletter
\patchcmd\longtable{\par}{\if@noskipsec\mbox{}\fi\par}{}{}
\makeatother
\IfFileExists{footnotehyper.sty}{\usepackage{footnotehyper}}{\usepackage{footnote}}
\makesavenoteenv{longtable}
\usepackage{graphicx}
\makeatletter
\def\maxwidth{\ifdim\Gin@nat@width>\linewidth\linewidth\else\Gin@nat@width\fi}
\def\maxheight{\ifdim\Gin@nat@height>\textheight\textheight\else\Gin@nat@height\fi}
\makeatother
\setkeys{Gin}{width=\maxwidth,height=\maxheight,keepaspectratio}
\makeatletter
\def\fps@figure{htbp}
\makeatother

\makeatletter
\@ifpackageloaded{caption}{}{\usepackage{caption}}
\AtBeginDocument{%
\ifdefined\contentsname
  \renewcommand*\contentsname{Table of contents}
\else
  \newcommand\contentsname{Table of contents}
\fi
\ifdefined\listfigurename
  \renewcommand*\listfigurename{List of Figures}
\else
  \newcommand\listfigurename{List of Figures}
\fi
\ifdefined\listtablename
  \renewcommand*\listtablename{List of Tables}
\else
  \newcommand\listtablename{List of Tables}
\fi
\ifdefined\figurename
  \renewcommand*\figurename{Figure}
\else
  \newcommand\figurename{Figure}
\fi
\ifdefined\tablename
  \renewcommand*\tablename{Table}
\else
  \newcommand\tablename{Table}
\fi
}
\@ifpackageloaded{float}{}{\usepackage{float}}
\floatstyle{ruled}
\@ifundefined{c@chapter}{\newfloat{codelisting}{h}{lop}}{\newfloat{codelisting}{h}{lop}[chapter]}
\floatname{codelisting}{Listing}

\makeatother
\makeatletter
\@ifpackageloaded{caption}{}{\usepackage{caption}}
\@ifpackageloaded{subcaption}{}{\usepackage{subcaption}}
\makeatother

\ifLuaTeX
  \usepackage{selnolig}  % disable illegal ligatures
\fi
\usepackage[]{natbib}
\usepackage{bookmark}

\IfFileExists{xurl.sty}{\usepackage{xurl}}{} % add URL line breaks if available
\hypersetup{
  pdftitle={Title},
  pdfauthor={Author 1; Author 2},
  pdfkeywords={3 to 6 keywords, that do not appear in the title},
  colorlinks=true,
  linkcolor={blue},
  filecolor={Maroon},
  citecolor={Blue},
  urlcolor={Blue},
  pdfcreator={LaTeX via pandoc}}

\newcommand{\anon}{1}

\usepackage{authblk}
\usepackage{pdflscape}
\usepackage{makecell}
\usepackage{threeparttablex}

\newtheorem{assumption}{Assumption}

\newcommand{\yes}{\checkmark}
\newcommand{\no}{$\times$}
\newcommand{\bone}{\mathbb{I}}
\newcommand{\ind}{\mbox{$\perp\!\!\!\perp$}}
\newcommand{\ntrial}{{n_1}}
\newcommand{\nec}{{n_0}}
\newcommand{\nt}{{n_{11}}}
\newcommand{\nc}{{n_{10}}}
\newcommand{\mupc}{{\mu_{0}}}

\usepackage{silence}
\usepackage{setspace}
\makeatletter
\renewcommand{\section}{\@startsection{section}{1}{0pt}
  {0.8pt}
  {0.4pt}
  {\normalfont\Large\bfseries}}

\renewcommand{\subsection}{\@startsection{subsection}{2}{0pt}
  {0.6pt}
  {0.3pt}
  {\normalfont\large\bfseries}}

\renewcommand{\subsubsection}{\@startsection{subsubsection}{3}{0pt}
  {0.5pt}
  {0.25pt}
  {\normalfont\normalsize\bfseries}}

\renewcommand{\paragraph}{\@startsection{paragraph}{4}{0pt}
  {0.3pt}
  {0.2pt}
  {\normalfont\normalsize\bfseries}}
\makeatother

\begin{document}

\def\spacingset#1{\renewcommand{\baselinestretch}%
{#1}\small\normalsize} \spacingset{1}

\if1\anon
{
  \title{\bf Externally Controlled Trials: A Review of Design and Borrowing Through a Causal Lens}

\author[1,2]{\normalsize Ke Zhu}

\author[3]{Rima Izem}

\author[4]{Peng Yang}

\author[4]{Ying Yuan}

\author[5,2]{Herbert Pang}

\author[6]{Mark van der Laan}

\author[7]{Lei Nie}

\author[8]{Birol Emir}

\author[7]{Pallavi Mishra-Kalyani}

\author[7]{Hana Lee}

\author[1]{Shu Yang\thanks{Address for correspondence: Shu Yang, Department of Statistics, North Carolina State University, Raleigh, NC, USA. Email: syang24@ncsu.edu}}

\affil[1]{\footnotesize Department of Statistics, North Carolina State University, Raleigh, NC, USA}

\affil[2]{Department of Biostatistics and Bioinformatics, Duke University School of Medicine, Durham, NC, USA}

\affil[3]{Statistical Methodology, Novartis Pharma AG, Basel, Switzerland}

\affil[4]{Department of Biostatistics, The University of Texas MD Anderson Cancer Center, Houston, TX, USA}

\affil[5]{PD Data Science \& Analytics, Genentech, South San Francisco, CA, USA}

\affil[6]{Division of Biostatistics, University of California, Berkeley, CA, USA}

\affil[7]{Office of Biostatistics, Center for Drug Evaluation and Research, Food and Drug Administration, Silver Spring, MD, USA\thanks{The views expressed in this manuscript are those of the author(s) and do not necessarily represent the official views of, nor an endorsement, by the U.S. FDA, HHS, or the U.S. Government.}}

\affil[8]{Pfizer Research \& Development, Chief Medical Office, Data Sciences and Analytics, New York, NY, USA}

\date{}
  \maketitle
} \fi

\if0\anon
{
\title{\bf Externally Controlled Trials: A Review of Design and Borrowing Through a Causal Lens}
\author{}
\date{}
\maketitle
} \fi

%\bigskip

\begin{abstract}
Externally controlled trials (ECTs) are increasingly used when randomized controls are infeasible, unethical, or insufficient, including applications in rare diseases, oncology, pediatrics, and post-approval effectiveness research. Although methodological work has expanded rapidly across causal inference, Bayesian dynamic borrowing, and hybrid trial designs, the literature remains fragmented. We adopt a six-step scientific roadmap to organize modern ECT methodology in two primary settings: (i) single-arm trials that evaluate efficacy through comparison with external controls, and (ii) hybrid controlled trials that augment the internal control arm with external controls drawn from real-world data or historical studies. The roadmap clarifies causal estimands, identifiability assumptions, and how statistical parameters arise from identification, and shows how modeling and borrowing strategies trade off efficiency and robustness, especially under covariate shift and outcome drift. Within this framework, we synthesize and evaluate recent Bayesian and frequentist developments, compare their strengths, limitations, operating characteristics, and available software, and emphasize the role of sensitivity analysis. By re-framing ECT methodology through a causal lens, this work establishes a coherent foundation for integrating external data into regulatory and clinical decision-making and highlights core challenges and opportunities for future research.
\end{abstract}

\noindent%
{\it Keywords:} Bayesian dynamic borrowing; causal inference; 
hybrid trial designs; real-world evidence; sensitivity analysis.
\vfill

\newpage
\spacingset{1.75} % DON'T change the spacing!

\section{Introduction}

Randomized controlled trials (RCTs) are widely regarded as the gold standard for evaluating the efficacy and safety of medical interventions. Their primary strength lies in the protection afforded by randomization: treatment groups are comparable prior to intervention, outcomes are measured under standardized protocols, and causal effects can be estimated with minimal reliance on modeling assumptions.

In contemporary settings, however, traditional RCTs can sometimes be difficult or impossible to conduct at the scale or within the time-frame required for regulatory decision-making. Recruitment challenges in rare diseases, ethical concerns about placebo controls in severe or rapidly progressing conditions, high operational costs, long follow-up requirements, and the need for timely post-approval evidence have all spurred interest in incorporating data external to the trial. 
These considerations have led to the growing use of external controls (ECs), consisting of patients drawn from historical clinical trials, disease registries, natural history studies, or real-world data (RWD) sources such as electronic health records and insurance claims. Such data may be available at the design stage, collected concurrently with the trial, or prospectively accrued alongside the clinical study.

The integration of ECs into clinical development gives rise to externally controlled trials (ECTs), a broad class of designs in which experimental treatment arms are compared, directly or indirectly, with patients not randomized within the same trial \citep{pocock1976combination,hampson2023innovative}, see Figure~\ref{fig:ECT}. Regulatory agencies and industry sponsors increasingly view ECTs as a
practical complement to conventional trials, particularly in marketing authorization in rare diseases and oncology \citep{fda2019rare,MHRA2025ExternalControlArms,EMA2025ExternalControlsWorkshop}. Beyond these settings, ECTs can inform industry decision making in early development, and benefit-risk assessments in pediatrics and post-market settings \citep{ICH-M14,FDA_CID_Meeting_Program}. 
Notably, ECTs share similar components with complex innovative designs such as master protocols and platform trials, where borrowing can occur across nonconcurrent arms, time periods, and disease subpopulations.
Despite these advantages and growing applications, the reliability of evidence from ECTs hinges on strong and often unverifiable assumptions about the relevance and quality of the external data and the analytic methods employed, making principled data selection, quality by design, and rigorous analysis essential.

Over the past decade, methodological research on ECTs has expanded rapidly \citep{rahman2021leveraging,mishra2022external,ventz2022design,polley2024leveraging,chen2024power,kotecha2024leveraging,urru2025integrating,ye2024considerations,ung2025keep,selukar2025synthetic}. Proposed approaches include causal inference methods adapted from observational studies, Bayesian dynamic borrowing, and emerging frequentist robust borrowing strategies such as prognostic adjustment, test-then-pool, selective borrowing, bias-model-based integration, estimator averaging, randomization inference, and sensitivity analysis. Despite this progress, methods are often presented in isolation with differing estimands, assumptions, and inferential goals, making it difficult for practitioners and regulators to relate approaches or choose methods for a given scientific question.

\textbf{Scope and contribution.} This article synthesizes and evaluates modern ECT methodology within a unified causal framework, clarifying the conceptual structure underlying different approaches, identifying the assumptions on which their validity depends, and offering a scholarly perspective on their relative strengths, limitations, and appropriate scope of application. We make three main contributions: first, we adopt a six-step scientific roadmap for ECTs, spanning estimand specification, observed data structure, identifiability assumptions, statistical identification, estimation, and sensitivity analysis, showing that many apparent differences between Bayesian and frequentist approaches stem primarily from distinct exchangeability assumptions; second, we systematically review two primary ECT designs, namely single-arm trials with external controls (SAT+EC) and hybrid controlled trials (HCT), comparing their efficiency-robustness trade-offs and available software implementations; third, we provide a forward-looking perspective on open challenges and emerging directions, including the role of ECT methodologies in master protocol and platform trial settings, the principled integration of AI-generated or synthetic data, and the integration of Bayesian and frequentist paradigms.

The article is organized as follows. Section~\ref{sec:notation} introduces two ECT designs and the scientific roadmap. Section~\ref{sec:sat} reviews methods in SAT with ECs, covering the estimand, data structure, identifiability, inference, and sensitivity analysis. Section~\ref{sec:hct} presents HCTs using the same framework and reviews recent Bayesian and frequentist borrowing methods. Section~\ref{sec:con} concludes with lessons learned, a discussion of controversies, open challenges, and opportunities for future research.

\section{Overview of ECT designs and a scientific roadmap}
\label{sec:notation}

\begin{figure}[t]
    \centering
    \includegraphics[width=\linewidth]{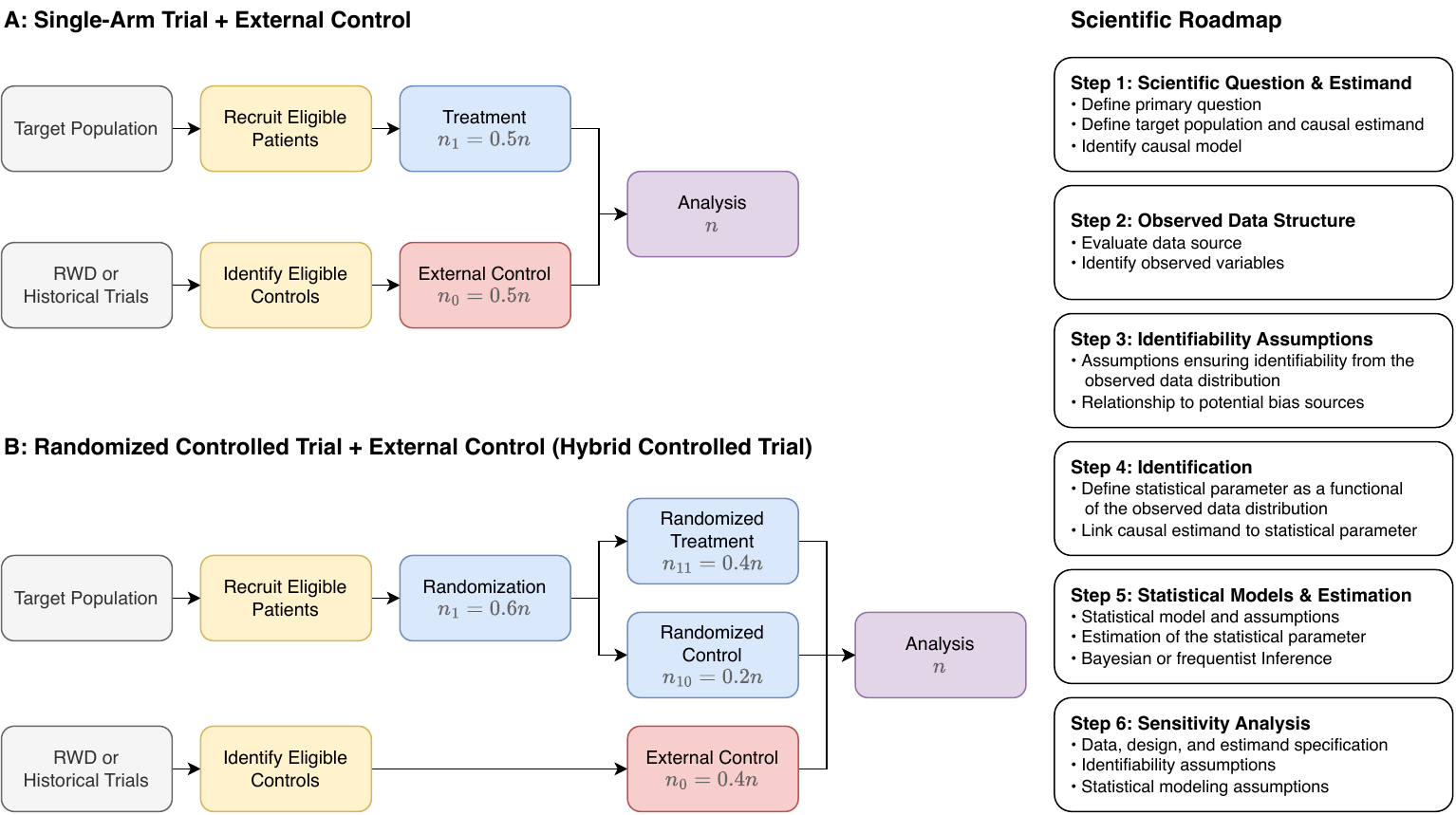}
    \caption{Two ECT designs and scientific roadmap. Gray assignment ratios are illustrative only and may vary by context.}
    \label{fig:ECT}
\end{figure}

This paper splits ECTs into two types (Figure~\ref{fig:ECT}) based on the presence of \textbf{randomization}, a key distinction with important statistical and practical implications that yields different strengths and limitations of each design type \citep{aronow2025nonparametric}.

\textbf{Single-arm trial evaluating efficacy via comparison with EC.} This design involves a single treatment arm trial \textbf{without randomization}, relying entirely on ECs for comparison (Figure~\ref{fig:ECT}, A). This design is relevant in contexts where randomization is unfeasible or unethical, as seen in rare disease or oncology settings.  The added value of this design is to enable evidence generation on comparative efficacy or safety of the investigative treatment in the single arm with other control treatments in the ECs. The challenges arise primarily from confounding bias in non-randomized comparisons. Addressing this bias requires untestable assumptions about between-source differences. We review SAT+EC in Sections~\ref{sec:sat}.  
    
\textbf{Hybrid controlled trial.} This design augments the RCT with ECs (Figure~\ref{fig:ECT}, B) to increase power for comparisons that may be underpowered in the RCT alone. The main challenge is that anticipated efficiency gains may be offset by bias and type I error arising from heterogeneity between data sources. HCTs are generally less prone to bias than SAT+EC designs because \textbf{randomization} ensures that (i) an RCT-only analysis provides a valid, though potentially underpowered, benchmark, and (ii) the RCT control group enables detection of bias in ECs due to measured and unmeasured confounding. We focus on settings where the RCT and ECs have comparable follow-up durations in Sections~\ref{sec:hct}, and discuss settings with longer EC follow-up in Section~\ref{sec:con}.

\begin{figure}[t]
    \centering
    \includegraphics[width=0.8\linewidth]{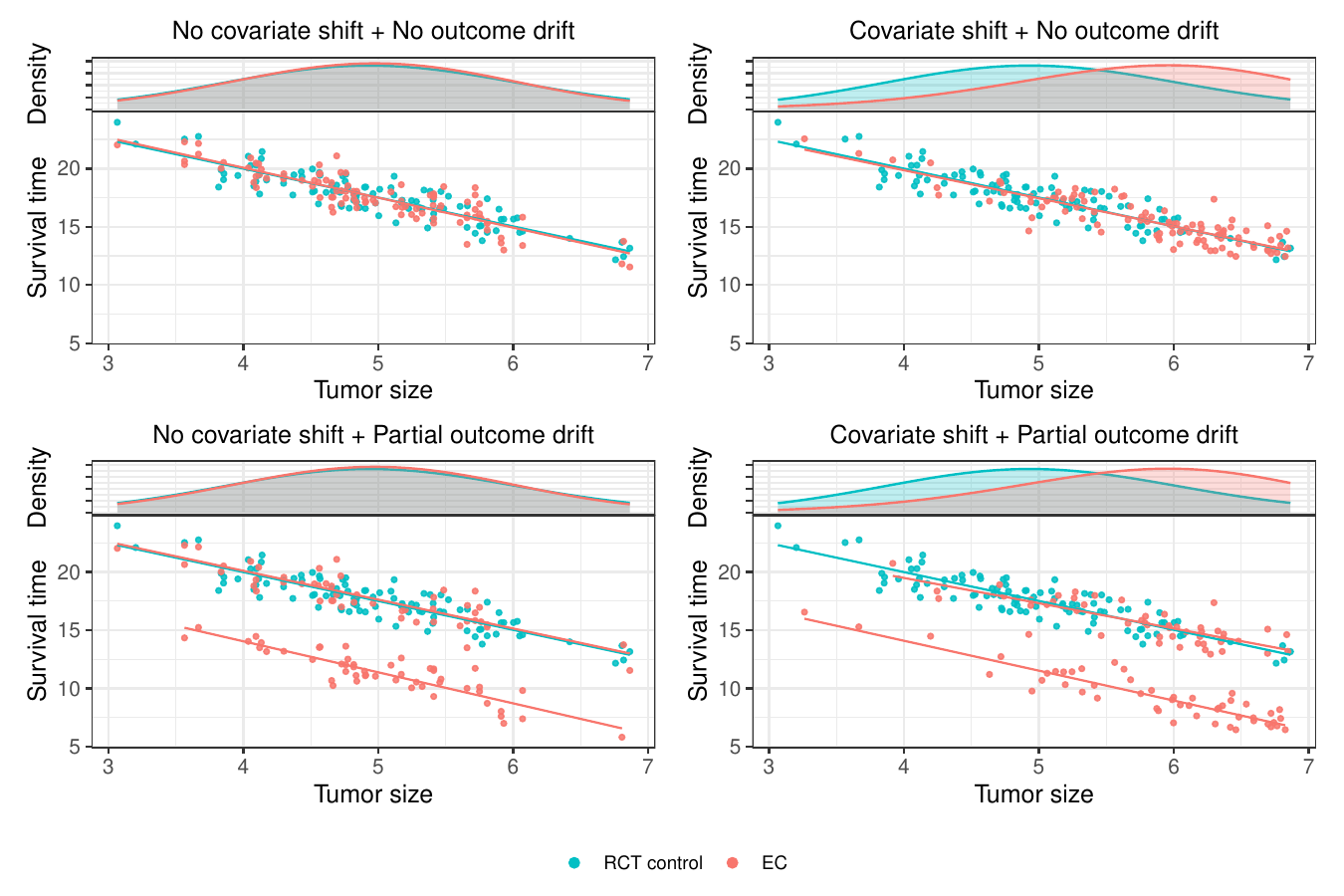}
    \caption{Illustration of covariate shift and outcome drift.}
    \label{fig:XshiftYdrift}
\end{figure}

The estimation of treatment effects under either design is prone to two types of bias: (i) \textbf{covariate shift}, defined as systematic differences in \textit{observed baseline covariates} between the trial population and ECs; and (ii) \textbf{outcome drift}, referring to residual differences in outcomes between trial controls and ECs after conditioning on comparable baseline covariates. 
A visual illustration is provided in Figure~\ref{fig:XshiftYdrift}, where the covariate is tumor size and the outcome is survival time.
This paper focuses on methods that address these biases and rely on accessible \textit{subject-level data} for ECs, including covariates, treatment assignment, and outcomes. In practice, EC sources are often much larger and more heterogeneous than the trial. As illustrated in Figure~\ref{fig:ECT}, a first step is to define a comparable EC cohort through \textit{eligibility alignment}, enabling the EC to align with the trial before downstream analysis.

\textbf{A scientific roadmap.} To navigate the varied landscape of ECT methodologies, we adopt the scientific roadmap \citep[e.g.,][]{petersen2014causal,van2011targeted,dang2023causal,gruber2023evaluating,dang2023start,hampson2024combining}. 
This roadmap structures study planning and evaluation through the six steps in Figure~\ref{fig:ECT}.

\begin{table}[t]
\centering
\caption{Unified notation for SAT+EC and HCT designs}
\label{tab:notation}
\small
\begin{tabular}{p{0.11\linewidth} p{0.85\linewidth}}
\toprule
\textbf{Symbol} & \textbf{Description} \\
\midrule

$X$ & Baseline covariates \\

$S \in \{1,0\}$ & Data source; $1$ = trial (SAT/RCT), $0$ = EC \\

$A \in \{1,0\}$ & Treatment; $1$ = treated, $0$ = control \\

$Y$ & Observed outcomes over a prespecified time frame, without time-varying treatment. \\

$Y(a)$ & Potential outcome under $a \in \{0,1\}$ \\

\midrule
\multicolumn{2}{l}{\textit{Propensity and outcome models}} \\

$\pi(x)$ & Sampling propensity score (typically unknown), $\mathbb{P}(S=1 \mid X=x)$ \\

$e(x)$ & Treatment propensity score in the trial (known by design), $\mathbb{P}(A=1 \mid S=1, X=x)$ \\

$\mu_{1a}(x)$ & Outcome model for arm $a$ in the trial, $\mathbb{E}[Y \mid S=1, A=a, X=x]$ \\

$\mupc(x)$ & Outcome model for pooled controls, $\mathbb{E}[Y \mid A=0, X=x]$ \\

\midrule
\multicolumn{2}{l}{\textit{Scientific and causal estimands}} \\

$\theta_a$ & Arm-specific mean outcome in the trial population, $\mathbb{E}[Y(a)\mid S=1]$ \\

$\tau$ & Average treatment effect (ATE) in the trial population, $\mathbb{E}[Y(1)-Y(0)\mid S=1]$ \\

\midrule
\multicolumn{2}{l}{\textit{Sample sizes}} \\

$n$ & Total sample size \\

$\ntrial$ & Trial sample size (SAT/RCT), $\sum_{i=1}^n \bone(S_i=1)$ \\

$\nt$ & Treated sample size in the trial, $\sum_{i=1}^n \bone(S_i=1, A_i=1)$ \\

$\nc$ & Control sample size in the trial, $\sum_{i=1}^n \bone(S_i=1, A_i=0)$ \\

$\nec$ & EC sample size, $\sum_{i=1}^n \bone(S_i=0)$ \\

\bottomrule
\end{tabular}
\end{table}

Table~\ref{tab:notation} gives the unified notation for both SAT+EC and HCT designs under the potential outcomes framework \citep{neyman1923application,rubin1974estimating}. We take the trial population as the target population and assume the sampling positivity condition $\mathbb{P}(S=1) > 0$ throughout.

\vspace{-5pt}

\section{Single-arm trials with external controls}
\label{sec:sat}

\subsection{Scientific question and estimand}
\label{sec:ec-estimand}

The scientific question concerns comparing the effect of a treatment versus control on a specific outcome for patients with a given indication. 
Different frameworks such as the target trial emulation \citep{hernan2016using}, the estimand framework \citep{ICH_E9_R1_2019} or a combination of the two frameworks \citep{hampson2024combining,polito2024applying}, can add more granularity to the scientific question. These frameworks translate the comparative objective into an estimand with specific attributes: population, initiated treatment and treatment strategies of interest (in the comparison), variable (outcome and follow-up time), intercurrent events of interest, and summary measure.

\textbf{Target population.}
The target population includes patients eligible for the target trial comparing the investigational treatment to control, defined by key eligibility criteria (e.g., protocol-specified medical criteria, region, or calendar time), typically a subset of those in the SAT.
Emulating the target trial requires aligning these sources with each other and with the target trial on key attributes \citep{FDA2023ExternallyControlledTrials}. This alignment should prioritize key prognostic factors identified in collaboration with clinical experts to minimize potential confounding bias in the comparison.

\textbf{Outcome for comparison, follow-up, and intercurrent events.}
The outcome quantifies how a patient feels or functions for a meaningful comparison of treatments. The outcome's definition and assessments should be aligned in the SAT and the EC for a valid comparison. They could differ from the primary outcome or time-frame of assessment in the SAT provided the comparison complements or adds value to the SAT findings. For example, meaningful definition of responder, and handling of intercurrent events of treatment discontinuation or use of rescue therapy may differ for the SAT and for the external data. 

\textbf{Causal estimand.}
The average treatment effect on the treated (ATT) is often the causal estimand in the SAT+EC setting when the target population is anchored at eligibility in the SAT. 
The ATT may coincide with the ATE if the EC emulates the control in the target trial without the need for additional eligibility criteria.

\subsection{Observed data structure}

At trial completion, the same subject-level dataset includes SAT and ECs. Table~\ref{tab:data-sat} (left) summarizes the data structure, where $S=A$ and $X$ are baseline covariates. We assume $\nec$ eligible units after eligibility alignment.
The data should be fit-for-purpose, with $Y$ and $X$ observed, reliable, and consistently defined across sources. Missingness, especially in $Y$, can compromise inference and should be reported as a data quality assessment. Analytic approaches for handling missingness require additional assumptions (e.g., missing at random) and evaluation of their plausibility. Variables $X$ should be identified and justified at the design stage using subject-matter expertise (e.g., literature review, directed acyclic graphs). They include key prognostic variables of $Y$ in the target population, such as disease severity, life-style, or healthcare access. Instrumental variables that are associated with $A$ but not prognostic of $Y$ should be avoided \citep{vanderweele2019principles}. Lastly, variables $X$ should exclude post-baseline variables that lie on the causal pathway between treatment and outcome.

\begin{table}[t]
\centering
\caption{Observed data structures in a SAT+EC (left) and a HCT (right). A check mark denotes an observed value. See Figure~\ref{fig:ECT} for a visual representation.}

\begin{minipage}[t]{0.48\textwidth}
\centering
\begin{tabular}{lcccc}
\toprule
Unit & $S$ & $A$ & $X$ & $Y$ \\
\midrule
1 & 1 & 1 & \checkmark & \checkmark \\
$\vdots$ & $\vdots$ & $\vdots$ & $\vdots$ & $\vdots$ \\
$n_{\text{1}}$ & 1 & 1 & \checkmark & \checkmark \\
$n_{\text{1}}+1$ & 0 & 0 & \checkmark & \checkmark \\
$\vdots$ & $\vdots$ & $\vdots$ & $\vdots$ & $\vdots$ \\
$n_{\text{1}} + \nec$ & 0 & 0 & \checkmark & \checkmark \\
\bottomrule
\end{tabular}
\end{minipage}
\hfill
\begin{minipage}[t]{0.48\textwidth}
\centering
\begin{tabular}{lcccc}
\toprule
Unit & $S$ & $A$ & $X$ & $Y$ \\
\midrule
1 & 1 & 1 & \checkmark & \checkmark \\
$\vdots$ & $\vdots$ & $\vdots$ & $\vdots$ & $\vdots$ \\
$\nt$ & 1 & 1 & \checkmark & \checkmark \\
$\nt+1$ & 1 & 0 & \checkmark & \checkmark \\
$\vdots$ & $\vdots$ & $\vdots$ & $\vdots$ & $\vdots$ \\
$\nt+\nc$ & 1 & 0 & \checkmark & \checkmark \\
$\nt+\nc+1$ & 0 & 0 & \checkmark & \checkmark \\
$\vdots$ & $\vdots$ & $\vdots$ & $\vdots$ & $\vdots$ \\
$\nt+\nc + \nec$ & 0 & 0 & \checkmark & \checkmark \\
\bottomrule
\end{tabular}
\end{minipage}

\label{tab:data-sat}
\end{table}

\vspace{-5pt}

\subsection{Identifiability assumptions}

\begin{assumption}\label{ass:sat}
(i) Consistency and no interference: $Y = A  Y(1) + (1 - A)  Y(0)$. (ii) No unmeasured confounders for treatment: $\{Y(1), Y(0)\} \ind A \mid X$. (iii) Positivity of treatment assignment: $\pi(x)<1$ for all $x$ such that $\mathbb{P}(X=x\mid A=1) > 0$.
\end{assumption}

Assumption~\ref{ass:sat}(i) requires a well-defined treatment with no multiple versions and no interference between units; this is plausible in drug studies when treatment and control are consistently defined across sites but may fail in settings where one individual's treatment can affect others' outcomes (e.g., infectious diseases). This assumption can also be violated in the presence of intercurrent events. For example, if treatment discontinuation occurs among those assigned to treatment, the observed outcome $Y$ may differ from the potential outcome $Y(1)$; similarly, if control patients receive rescue medication, their observed outcome may differ from $Y(0)$. Assumption~\ref{ass:sat}(ii) requires that all confounders of the treatment–outcome relationship are measured in both the trial and external data. Assumption~\ref{ass:sat}(iii) requires overlap in the covariate distribution among treated units; its plausibility can be assessed through covariate balance and overlap in estimated propensity score distributions \citep{greifer2020covariate}. In practice, Assumptions~\ref{ass:sat}(ii)–(iii) are often vulnerable when combining SAT and EC sources because between source differences in latent factors such as concurrency, measurement error, or follow-up intensity can introduce confounding. In such cases conditioning on $X$ may not ensure conditional independence or overlap, motivating the sensitivity analyses in Section~\ref{sec:SAT-sen}. Assumption~\ref{ass:sat} is also required for identification of the causal estimand with Bayesian approaches \citep{li2023bayesian,muller2023bayesian}.

\subsection{Identification}

Under Assumption~\ref{ass:sat}, the causal estimand $\tau$ is identifiable  using the data from the SAT + EC design. This relies on viewing the data as an observational study with treatment assignment governed by the sampling mechanism. As a result, the causal estimand is identified as
$\tau=\mathbb{E}[Y\mid A=1]-\mathbb{E}\!\left\{
\mathbb{E}[Y\mid A=0,X]\mid A=1
\right\}$,
which can be further expressed as a functional of the observed data distribution via outcome modeling, weighting by the sampling propensity score $\pi(X)$, or doubly robust combinations thereof. This identification strategy parallels that of standard observational studies targeting the ATT.

\subsection{Statistical models, estimation, and inference}
Given the identifiability assumptions in Step~4, estimation in the SAT+EC setting reduces to a standard causal inference problem targeting the ATT. Consequently, a wide range of methods developed for observational studies are relevant \citep{imbens2015causal,hernan2020causal}. 
However, simulations to evaluate the finite sample size performance and reliability of these methods is warranted when the SAT+EC is the primary basis of efficacy in small population settings.
We briefly review these methods through two interconnected components: (i) statistical design and (ii) treatment effect estimation and inference in Supplementary Material Section~B.

\subsection{Sensitivity analyses}
\label{sec:SAT-sen}

We focus on two types of sensitivity analyses: (i) data, design, and estimation specification; (ii) identifiability assumptions. A key principle in planning is \textit{prespecification} of these analyses in the statistical analysis plan to mitigate post hoc rationalization.

\textbf{Sensitivity to data, design, and estimation specifications.} The design process of SAT+EC studies includes multiple choices for specifications such as eligibility alignment, outcome specification, follow-up window, handling of intercurrent events, or choice of summary measure. Also, results may depend on implementation decisions in Step~5, including model specifications for $\pi$ and $\mu_a$, or restrictions on those models \citep{yu2020veridical}. Sensitivity analyses can examine the robustness of findings across closely related and scientifically plausible alternative design or analytic specification choices.

\textbf{Sensitivity to violations of identifiability assumptions.} Identification of $\tau$ relies on the identifying assumptions above, that are untestable with observed data. A simple bias formulation introduces a \emph{causal gap} parameter $\delta$ representing residual bias, defining $\tilde{\tau}(\delta) = \hat{\tau} - \delta$, where $\hat{\tau}$ is the estimator from Step~5 \citep{diaz2013sensitivity}. Varying $\delta$ over a prespecified range shifts the confidence interval for $\hat{\tau}$; the \textit{tipping point} is the smallest $|\delta|$ for which the shifted interval includes zero. This formulation clarifies the tradeoff among effect size, standard error, and the magnitude of unmeasured bias needed to overturn the conclusion. If a small $\delta$ alters inference, conclusions are fragile; if a large $\delta$ is required, findings are more robust. Multiple methods exist for assessing the extent or impact unmeasured confounding, the \textit{E-values} \citep{vanderweele2017sensitivity,mathur2018web} quantify the minimum strength of association an unmeasured confounder must have with both treatment and outcome to explain away the observed effect. Also, negative control outcomes or exposures, when available, can probe the unmeasured confounding \citep{song2026proximal}. The two-level sensitivity framework by \cite{gao2025doubly} addresses both unmeasured confounding and intercurrent events. The review in \cite{fang2025sensitivity} and the practical guidance for real-world evidence studies \citep{faries2025real} include additional considerations. These approaches illustrate quantitative assessment of impact of departure from an assumption rather than its validation.

\subsection{Software}
Implementations of causal inference methods in R is discussed in the CRAN task view \citep{mayer2022cran}. The website \url{https://www.elizabethstuart.org/psoftware} includes a comprehensive list of libraries implementing propensity score methods. TMLE implementations are provided through the \texttt{tlverse} software ecosystem \citep{vdl2023tlverse}. \href{https://github.com/Gaochenyin/SensDR}{\texttt{SensDR}} provides a dual sensitivity analysis, assessing robustness to violations of no unmeasured confounding and the presence of intercurrent events.

\section{Hybrid controlled trials}
\label{sec:hct}

\subsection{Scientific question and estimand}
\label{sec:sci_hct}

This paper focuses on comparative scientific questions of the treatment to control for a specific outcome observed within the same time frame, and for the same target population of those eligible to the RCT. 
The choice of the RCT population $(S=1)$ as an anchor for the inference is sensible when answering the scientific objective prioritizes internal validity for the comparison of treatments, often of primary interest in pharmaceutical research and regulatory assessment during development and at marketing approval decisions. Alternatives to this choice include scientific questions aiming to extrapolate, transport, or generalize the treatment effect beyond the clinical trial population or the time-frame of observation in the RCT. While those questions are relevant in public health, they are beyond the scope of our paper as they call for different identifying assumptions and methods of transportability, generalizability, or assessment of treatment effect heterogeneity that we discuss in Supplementary Material Section C.   

The focus on a comparison between treatment and control, the same as in the RCT also prioritizes the control selected in the RCT over other potential sources in the primary objective.  
For simplicity, we consider variables or outcomes $Y$ to be a continuous endpoint determined at a particular time-point and summary measures of differences in means. These could be the same variable as in one of the objective estimands of the RCT alone or could be extended to other variables that are potentially underpowered with the RCT alone. Note that the set of estimands and the process can be extended to longitudinal outcomes (e.g., \cite{zhou2024causal}) or to time-to-event data. For example, causal cumulative incidence has been discussed in \cite{su2025improving}, and the win ratio in \cite{chen2026propensity}. We also refer to \cite{colnet2023risk} and \cite{boughdiri2025unified} for more causal measures.

\subsection{Observed data structure}

At study completion, the analytical dataset includes all those in the RCT and a relevant subset of those in the EC. Table~\ref{tab:data-sat} (right) summarizes the observed data structure in a HCT. 
This paper focuses on HCTs where only the control group data is augmented by the EC. That is because in clinical development, data on an investigative treatment is often more scarce than on an existing standard of care. Also, augmenting controls can allow fewer patients to be randomized into the RCT control arm. 

As discussed with the previous design, the EC analytical dataset could be a subset of those meeting eligibility criteria in the trial to ensure comparability of sources and start of follow-up by design. Target trial emulation considerations can systematically assess and ensure comparability of follow-up, and outcome definition. Comparability of data capture of prognostic variables $X$ across data sources is also critical.

When external data include both control and treatment arms \citep{colnet2024causal,dang2022cross,van2024adaptive,parikh2025double,rosenman2025methods}, we can borrow from both arms and multiple trials or studies. Nevertheless, they would require additional identifying assumptions and considerations of exchangeability across studies. We discuss additional data structures in Supplementary Material Section~C.

\subsection{Identifiability assumptions}
\label{sec:iden_hct}

\subsubsection{Assumption for RCT}

To identify the causal estimand, we adopt standard RCT identifiability assumptions under which $\tau$ is identifiable from RCT data alone and typically ensured by randomization.

\begin{assumption}[RCT identification]\label{ass:rct}
(i) Consistency and no interference: $Y = A  Y(1) + (1 - A)  Y(0)$. (ii) No unmeasured confounders for treatment: $\{Y(1), Y(0)\} \ind A \mid (X, S = 1)$. (iii) Positivity of treatment assignment: $0 < e(x) < 1$ for all $x$ such that $f_{X \mid S}(x \mid s = 1) > 0$.
\end{assumption}

\subsubsection{Assumption for EC}

To improve the efficiency of estimating $\theta_0$, we may borrow EC data. However, doing so requires further assumptions to ensure sufficient homogeneity between the EC and RCT. 

\begin{assumption}[Conditional mean exchangeability]\label{ass:y}
$
\mathbb{E}[Y(0) \mid S = 1, X] = \mathbb{E}[Y(0) \mid S = 0, X].
$
\end{assumption}

\begin{assumption}[Positivity of sampling]\label{ass:pi}
$\pi(x)>0$ for all $x$ such that $f_{X}(x) > 0$.
\end{assumption}

Assumption~\ref{ass:y} holds if any discrepancy between the RCT controls and ECs is fully explained by measured confounders. It is weaker than sampling unconfoundedness, $Y(0)\ind S\mid X$, which requires the full conditional distribution of $Y(0)$ to coincide across sources given $X$. 
Violations may arise from unmeasured confounding or other factors (Supplementary Material Section~A), consequently, bias may remain after adjusting for observed covariates. Two strategies address this issue: (i) developing methods that relax Assumption~\ref{ass:y} (Sections~\ref{sec:bayes} and \ref{sec:freq}), or (ii) evaluating its plausibility through sensitivity analyses (Section~\ref{sec:hct-sen}).
Assumption~\ref{ass:pi} could be satisfied by restricting ECs to those that meet the eligibility criteria.

\subsection{Identification}

Under the identifiability assumptions in Section~\ref{sec:iden_hct}, the scientific estimands $\theta_a$ can be expressed as statistical parameters, i.e., functionals of the observed data distribution. Because different identifiability assumptions imply different links between the scientific estimand and the statistical parameter, we consider the two scenarios separately.

\subsubsection{Identification under Assumption \ref{ass:rct} (RCT-only)}
\label{sec:rctid}

Under Assumption~\ref{ass:rct}, the scientific estimand is identifiable using RCT data alone. 
Note that we consider the setting where both RCT and EC data are available with total sample size $n$, although only the RCT data are used for identification. Therefore, the identification formula differs from that in the setting where only RCT data are available, although both lead to the same estimator.
For $a\in\{0,1\}$,
the outcome model (OM) representation is 
$\theta_a=\mathbb{E}[\mu_{1a}(X)\mid S=1]$.
The inverse probability weighting (IPW) representation is 
$
\theta_a = \mathbb{E}\!\left[ 
\frac{\bone(A = a)}{\mathbb{P}(A=a\mid X,S = 1)}Y\mid S=1
\right].
$
The augmented IPW (AIPW) representation is 
$
\theta_a = 
\mathbb{E}\!\left[
\mu_{1a}(X)+
\frac{\bone(A=a)}{\mathbb{P}(A=a\mid X,S=1)}\{Y - \mu_{1a}(X)\} \mid S=1
\right], 
$
which is the solution to the equation obtained by setting the expectation of the efficient influence function (EIF) to zero.
The EIF characterizes the semiparametric efficiency bound for regular asymptotically linear estimators of $\theta_a$ \citep{tsiatis2006semiparametric}. Efficient estimators can be obtained by solving EIF-based estimating equations to obtain doubly robust estimates (e.g., AIPW) or via TMLE, which achieves the same efficiency while preserving a plug-in form through targeted updates \citep{van2011targeted}.

\subsubsection{Identification under Assumptions \ref{ass:rct}, \ref{ass:y} and \ref{ass:pi} (no outcome drift)}
\label{sec:ecid}

Under Assumptions~\ref{ass:rct} and \ref{ass:y}, $\theta_0$ can be identified by incorporating EC data \citep{li2023improving,Valancius2024}. The OM representation is 
$\theta_0=\mathbb{E}[\mupc(X)\mid S=1]$.
The IPW representation is 
$
\theta_0 =
\frac{1}{\mathbb{E}[\pi(X)]}
\mathbb{E}\!\left[ \pi(X)
\frac{(1-A)}{\{1-e(X)\}\pi(X)+\{1-\pi(X)\}}\,Y
\right].
$
Alternative IPW representations also exist \citep{wang2025integrating}; for example,
$
\theta_0 =
w\mathbb{E}\left[\frac{1-A}{1-e(X)}Y\mid S=1\right]+
(1-w)\frac{1}{\mathbb{E}[\pi(X)]}
\mathbb{E}\!\left[ \pi(X)
\frac{1-S}{1-\pi(X)}\,Y
\right],
$
$w\in(0,1).$
The AIPW representation is 
$
\theta_0 = 
\frac{1}{\mathbb{E}[\pi(X)]}
\mathbb{E}\!\left[
S\,\mu_{0}(X)+
\pi(X)\frac{(1-A)S+(1-S)r(X)}{\{1-e(X)\}\pi(X)+\{1-\pi(X)\}r(X)}
\{Y - \mu_{0}(X)\}
\right],
$
where $r(x)=\mathbb{V}(Y\mid S=1,A=0,X=x)/\mathbb{V}(Y\mid S=0,A=0,X=x)$ is the conditional variance ratio between RCT controls and ECs used for inverse-variance weighting. Similar identifications apply to $\theta_1$ when an external treatment arm is available; otherwise, identification for $\theta_1$ can rely on the RCT-only formulation in Section~\ref{sec:rctid}.

\subsection{Statistical models, estimation, and inference}
\label{sec:hct-step5}

\textbf{Taxonomy and a summary table.}
A central challenge in integrating ECs with RCTs is the discrepancy between the two data sources, including covariate shift in baseline covariates and outcome drift, where outcomes differ across sources even after conditioning on covariates. Adjusting only for covariate shift can yield larger efficiency gains when outcome drift is absent but may introduce bias otherwise. Methods that account for outcome drift provide more robust inference but typically reduce efficiency gains, reflecting a bias--variance tradeoff, with Bayesian and frequentist methods offering different degrees of this tradeoff.
We classify existing approaches into four categories based on data usage and identifying assumptions: (i) RCT-only analysis under Assumption~\ref{ass:rct}; (ii) borrowing under Assumptions \ref{ass:rct}, \ref{ass:y} and \ref{ass:pi} when no outcome drift is present; (iii) borrowing under outcome drift via Bayesian methods; and (iv) borrowing under outcome drift via frequentist methods. Within each category there are multiple approaches. The Bayesian–frequentist categorization is not strict, as several recent methods integrate ideas from both paradigms. Table~\ref{tab:hct-sum} provides a concise summary that guides readers to the corresponding subsections.

\subsubsection{RCT-only analysis under Assumption \ref{ass:rct}}

Before introducing methods that borrow EC data, we first establish what can be achieved using only RCT data, which serves as the no-borrowing benchmark. A simple estimator is the no borrowing difference-in-means $\hat{\tau}^{\text{NB-DIM}}=\nt^{-1}\sum_{i=1}^n S_iA_iY_i-\nc^{-1}\sum_{i=1}^n S_i(1-A_i)Y_i$, though efficiency can often be improved through covariate adjustment \citep{FDA2023CovariateAdjustment}. One possible RCT-only covariate-adjusted estimator $\hat{\tau}^{\text{NB-AIPW}}$ is
\begin{equation}
\label{eq:rctom} 
\frac{1}{\ntrial} \sum_{i=1}^n S_i\left\{
 \hat{\mu}_{11}(X_i) 
 +\frac{A_i}{e(X_i)}\{Y_i-\hat{\mu}_{11}(X_i)\}
-\hat{\mu}_{10}(X_i)
-\frac{1-A_i}{1-e(X_i)}\{Y_i-\hat{\mu}_{10}(X_i)\}
\right\},
\end{equation}
where $\hat{\mu}_{1a}(x)$ is the fitted outcome model for arm $a$ using only RCT data. 
Recall that $\ntrial=\sum_{i=1}^nS_i$ is the trial sample size.
Since the treatment assignment probability $e(x)$ within RCT is determined by design and correctly specified, the doubly robust estimator above is always consistent even if the outcome model is misspecified \citep{tsiatis2008covariate,rosenblum2010simple,guo2023generalized,bannick2025general}.
The role of $\hat{\mu}_{10}(x)$ is to improve the efficiency of the difference-in-means estimator. This formula is agnostic to the functional form of the outcome model given covariates and to the data used to inform the outcome model. For example, the outcome model could be informed by prior literature, estimated from the RCT data, or estimated from EC data as discussed in Section~\ref{sec:prog}. The functional form could also incorporate underlying biological mechanisms, parametric linear regression, or data-driven ``black-box'' models using machine learning algorithms. When flexible machine learning methods replace parametric models, valid inference can be preserved via cross-fitting and TMLE \citep{van2011targeted}.

\subsubsection{Borrowing under Assumptions \ref{ass:rct}, \ref{ass:y} and \ref{ass:pi} (no outcome drift)}
\label{sec:ps}

In this section, we synthesize and evaluate methods that address covariate shift between $X \mid S=1$ and $X \mid S=0$ by treating the sampling indicator $S$ as the ``treatment'' in an observational study and applying causal inference methods to balance covariates across data sources. Using these methods alone requires Assumption~\ref{ass:y} to rule out outcome drift in ECs. We also briefly mention their combination with methods that address outcome drift in Sections~\ref{sec:bayes} and \ref{sec:freq}, but leave a detailed discussion to those two sections.

\paragraph{Sampling propensity score matching}
\label{sec:mat}
A common approach to address covariate shift is to match on the estimated sampling propensity score $\pi(x)$  \citep{rosenman2022propensity,liu2022matching}, leveraging tools from observational causal inference \citep{ho2011matchit}. 
Matching by $\pi(X)$ decreases covariate shift between sources before further modeling \citep{shan2022simulation}. In practice, this shift reduction preprocessing step could reduce the analytical set by eliminating unmatched EC and RCT units at the tails of the $\pi$ distribution. Because this method uses only a function of baseline covariates $X$ and the trial is randomized, the EC can be matched to either the full RCT sample, or to each RCT arm \citep{lin2023matching}. These targets are asymptotically equivalent under randomization but may differ in small samples.

\paragraph{Sampling propensity score weighting}
Weighting adjusts EC units so their covariate distribution resembles that of the RCT population and typically makes use of all the data, downweighting dissimilar units in the EC.  
From the identification formula in Section~\ref{sec:ecid}, an IPW estimator is
$
\hat{\theta}_0^{\mathrm{FB\text{-}IPW}} 
= \frac{1}{\ntrial}\sum_{i=1}^n 
\hat{\pi}(X_i)\frac{(1-A_i)}{\{1-e(X_i)\}\hat{\pi}(X_i)+\{1-\hat{\pi}(X_i)\}}\,Y_i,
$
where ``FB'' denotes full borrowing of EC units, $e(x)$ is known by design in the RCT, and $\hat{\pi}(x)$ estimates the sampling propensity score \citep{Valancius2024}.  
Alternative IPW estimator is
$
\hat{\theta}_0^{\mathrm{FB\text{-}IPW2}}  =
\frac{w}{\ntrial}\sum_{i=1}^n 
\frac{S_i(1-A_i)}{1-e(X_i)}\,Y_i+
\frac{1-w}{\ntrial}\sum_{i=1}^n 
\frac{(1-S_i)\hat{\pi}(X_i)}{1-\hat{\pi}(X_i)}\,Y_i,
$
$w\in(0,1)$.
When $\hat{\pi}(x)$ is correctly specified, the IPW estimator is consistent and asymptotically normal.  
While weighting uses all the data and addresses covariate shift, bias from factors such as outcome drift may still remain.
To address outcome drift, recent works combine propensity score matching or weighting with informative prior
\citep{lin2018propensity,lin2019propensity,chen2020propensity,liu2021propensity,lu2022propensity,yu2022power,lin2022incorporating,wang2022propensity,harton2023combining,wang2024propensity,qian2025matching,zhao2025ps}, multi-source exchangeability models \citep{wei2024propensity}, or adaptive TMLE \citep{Qiu2025}, see more discussion in Sections~\ref{sec:bayes} and \ref{sec:freq}.

\paragraph{Outcome modeling}
\label{sec:gcomp}
Outcome modeling methods build on the identification formula in Section~\ref{sec:ecid} by pooling RCT controls and ECs to estimate the control outcome model, also known as g-computation \citep{zhang2025outcome}. The OM estimator is
$
\hat{\theta}_0^{\mathrm{FB\text{-}OM}}
= {\ntrial}^{-1}\sum_{i=1}^n S_i\,\hat{\mu}_0(X_i),
$
where $\hat{\mu}_0(x)$ is fitted by pooled controls. $\hat{\theta}_0^{\mathrm{FB\text{-}OM}}$ averages predictions over the RCT covariate distribution, addressing covariate shift and potentially improving efficiency by relying on an outcome model fitted on pooled data. 
Consistency of $\hat{\theta}_0^{\mathrm{FB\text{-}OM}}$ requires correct specification of the outcome model and no outcome drift.
We refer to Sections~\ref{sec:bayes} and \ref{sec:bm} for Bayesian and frequentist approaches that further address outcome drift within the outcome model.

\paragraph{Doubly robust methods}
Building on the identification formula in Section~\ref{sec:ecid}, the AIPW estimator is 
$
\hat{\theta}_0^{\mathrm{FB\text{-}AIPW}} 
= 
\frac{1}{\ntrial}
\sum_{i=1}^n\left[
S_i\,\hat{\mu}_{0}(X_i)+
\hat{\pi}(X_i)\,
\frac{(1-A_i)S_i+(1-S_i)\hat{r}(X_i)}{\{1-e(X_i)\}\hat{\pi}(X_i)+\{1-\hat{\pi}(X_i)\}\hat{r}(X_i)}
\{Y_i - \hat{\mu}_{0}(X_i)\}
\right],
$
where $\hat{r}(x)$ estimates the conditional variance ratio of $Y(0)$ between RCT and EC controls \citep{li2023improving}. Under Assumption~\ref{ass:y}, the estimator is consistent if either the sampling model or the outcome model is correctly specified and attains the semiparametric efficiency bound when both are consistent. \cite{gao2025improving} propose augmented calibration weighting (ACW) to stabilize weights through balance-constrained optimization. 
\cite{Valancius2024} uses machine learning approaches for the nuisance models and also compares them to TMLE, which includes a targeting step to obtain an efficient and doubly robust estimator. In contrast to AIPW, TMLE also belongs to the g-computation class as in Section~\ref{sec:gcomp} and preserves the parameter space, thus providing better finite-sample performance. 
\citet{dai2025incorporating} introduce a doubly safe estimator for high-dimensional covariates that maintains efficiency gains under partial misspecification. 
Doubly robust estimators may exhibit finite-sample bias due to high-dimensional nuisance estimation, which can be mitigated by HAL, undersmoothing, and adaptively exploiting low-dimensional structure in the outcome regression \citep{phillips2025commentary}.
Despite these advantages, doubly robust methods do not address residual outcome drift beyond that explained by covariates.

\subsubsection{Borrowing under outcome drift: Bayesian approach}
\label{sec:bayes}

In the Bayesian paradigm, inference is based on the joint posterior distribution of $\theta_1$ and $\theta_0$, obtained by assigning priors to these parameters, for example, a non-informative prior for $\theta_1$ and an informative prior for $\theta_0$ to leverage EC information in the HCT setting. The treatment effect is then derived from this joint posterior distribution; for the ATE, its posterior mean is the difference between the posterior means of $\theta_1$ and $\theta_0$. Any nuisance parameters involved in the likelihood may be assigned non-informative priors and integrated out under the Bayesian framework.

\paragraph{Informative prior} The Bayesian paradigm provides a coherent framework to integrate EC data by constructing an informative prior from EC data as follows:
\vspace{-20pt}
\begin{equation}\label{fullinforprior}
    p(\theta_0 \mid D_{EC}) \propto L(D_{EC} \mid \theta_0 ) p_0(\theta_0).
    \vspace{-20pt}
\end{equation}
where $D_{EC}$ denotes the EC data, $L(\cdot)$ denotes the likelihood and $p_0(\theta_0)$ denotes a non-informative prior. Inference for $\theta_0$ is then made based on its posterior distribution after updating this prior with the RCT control data.  When multiple EC datasets are available, the meta-analytic predictive (MAP) prior \citep{neuenschwander2010summarizing} can be used. This approach employs a Bayesian hierarchical meta-analysis to synthesize information across EC datasets while accounting for between-study heterogeneity. These priors are fully informative in the sense that they incorporate all EC information into the RCT inference.

However, when outcomes from the RCT control arm and the EC data differ systematically, such as in the presence of unmeasured confounding, the use of fully informative prior may introduce bias. To address this issue, more sophisticated forms of informative priors have been proposed using different statistical strategies, including discounting EC information (e.g., power priors, elastic priors), shrinkage (e.g., commensurate priors), and mixing with a noninformative prior (e.g., mixture priors), as described below. 

\textbf{Power priors} \citep{chen2000power} address potential incompatibility between RCT and EC data by discounting the EC likelihood through a power parameter $a_0 \in [0,1]$, that is, $p(\theta_0 \mid D_{EC}) \propto L(D_{EC} \mid \theta_0 )^{a_0} p_0(\theta_0)$.  However, because compatibility between the RCT controls and the EC is rarely known {\it a priori}, pre-specifying  $a_0$ is challenging. Under the Bayesian paradigm, one can assign  $a_0$ a prior (e.g., a Beta prior) and allow the data to determine the amount of information borrowed from the EC, using the normalized power prior (NPP) for proper posterior inference \citep{duan2006evaluating}. Also, \citet{shen2023optimal} developed optimal NPPs that encourage borrowing when the EC and RCT data are compatible and discourage it otherwise. Further, \citet{shen2024anpp} established connections between NPPs and Bayesian hierarchical models (BHMs; see Section~\ref{sec:Bayesian_shrinkage}) and extended the approach to accommodate multiple historical datasets. Unfortunately, simply assigning a Beta prior to $a_0$ and relying on the data to determine the degree of borrowing does not yield satisfactory adaptation, especially in compatible cases, as the data provide very limited information for estimating $a_0$ \citep{neuenschwander2009note, pawel2023normalized}.

\textbf{Empirical Bayesian approaches} have therefore been proposed to improve the adaptivity of the power prior, enabling more appropriate borrowing of information from EC. For example, \citet{gravestock2017adaptive} proposed selecting $a_0$ to maximize the marginal likelihood; \citet{liu2018dynamic} used a p-value–based compatibility test; \citet{bennett2021novel} used equivalence-based tail probabilities; and \citet{pan2017calibrated} proposed the calibrated power prior, selecting $a_0$ based on a predefined congruence measure between RCT and EC data and calibrating it via simulation to ensure desirable operating characteristics, such as type I error control and consistency. A similar alternative by \citet{lin2025combining} proposed a power likelihood approach, in which the power parameter is selected by maximizing the expected log predictive density to estimate the CATE given a targeted covariate of interest. Also, \citet{kwiatkowski2024case} proposed a case-weighted power prior using a piecewise proportional hazards model with discounting parameters on each pre-specified time interval, where compatibility between RCT and EC data is assessed via the posterior predictive distribution. 

\textbf{Elastic priors} \citep{jiang2023elastic} also use a discounting strategy to address potential incompatibility between RCT and EC data, but they differ from power priors in two important ways. First, the approach is more direct: it discounts the fully informative prior \eqref{fullinforprior} by inflating its variance, rather than discounting the likelihood. Second, the discount is data-driven, depending directly on the degree of incompatibility between the RCT and EC data through an empirical Bayes procedure. The core concept involves calibrating an elastic function $g(T)$, which maps a congruence measure $T$, quantifying the compatibility between RCT and EC data, to the interval $(0, 1)$. The elastic prior is then constructed by inflating the variance of the prior $p(\theta_0 \mid D_{EC})$ by a factor of $g(T)^{-1}$, thereby adaptively discounting EC information in accordance with the strength of compatibility. A key advantage of the elastic prior is that it incorporates clinical considerations when calibrating the elastic function, enabling more precise adaptive information borrowing, which in turn improves type I error control or power. In addition, the elastic prior satisfies information-borrowing consistency, that is, it borrows fully when the RCT and EC data are fully compatible or exchangeable, and refrains from borrowing when they are not, provided that the number of observations (rather than the number of trials) in the RCT and EC data is sufficiently large. This desirable property may not hold for the power prior. In contrast, achieving information-borrowing consistency for methods such as the commensurate prior (described below) often requires a large number of studies. For exponential family models, discounting the EC likelihood $L(D_{EC} \mid \theta_0 )$ by $a_0$ is equivalent to inflating the variance of the fully informative prior \eqref{fullinforprior} by $1/a_0$. Thus, many of the aforementioned empirical Bayes modifications of the power prior, such as the calibrated power prior, can be viewed as special cases of the elastic prior, corresponding to specific choices of the elastic function and congruence measure. This provides a link between elastic priors and power priors.

Unlike the power and elastic priors, that control borrowing by discounting the information from EC data, the \textbf{commensurate prior} \citep{hobbs2011hierarchical} adopts a different Bayesian strategy, shrinkage, to regulate the degree of information borrowing. Specifically, given EC data $D_{EC}$, the commensurate prior is 
$p(\theta_0 \mid D_{EC}, \theta_0^{EC} , \eta) \propto L(D_{EC} \mid \theta_0^{EC} ) p(\theta_0 \mid \theta_0^{EC}, \eta) p_0(\theta_0),
$
which induces shrinkage of the RCT control parameter, $\theta_0$, toward its counterpart from the EC data, $\theta_0^{EC}=\mathbb{E}[Y\mid S=0]$, through a commensurability (shrinkage) parameter $\eta$. The parameter $\eta$ represents the between-study variance, that is, the variability between the RCT and EC data, and therefore determines the extent of shrinkage toward the EC mean. \citet{hobbs2012commensurate} extended this framework to accommodate multiple historical datasets and generalized linear models.
Conceptually, the commensurate prior can be viewed as a special case of a BHM, where the first level specifies the likelihood for the RCT and EC data, and the second level shrinks the corresponding mean parameters toward each other through the prior  $p(\theta_0 \mid \theta_0^{EC}, \eta)$.
The commensurate prior faces two major challenges: (i) The estimation of $\eta$ is intrinsically difficult because it is informed by the number of datasets (e.g., just RCT and EC) rather than the number of subjects, leading to unstable borrowing decisions. In practice, one can use a proxy for studies by partitioning a large dataset into substudies. However, this approach may underestimate true between-study heterogeneity. This issue is well recognized in the meta-analysis literature, given the BHM representation of the commensurate prior. Consequently, its operating characteristics are highly sensitive to the prior specification of $\eta$. (ii) The approach requires integration over nuisance parameters, which can be computationally intensive and less scalable. This integration is somewhat redundant because the commensurate prior is equivalent to a BHM, and it is more straightforward and transparent to use BHM.

Another approach to address potential incompatibility between RCT and EC data is to use a mixture of an informative prior and a noninformative prior, where the mixing weight controls the amount of information borrowed from the EC data. \citet{schmidli2014robust} proposed the \textbf{robust MAP (rMAP) prior}, 
$
p(\theta_0 \mid D_{EC}) = w p {_{\text{MAP}}}(\theta_0 \mid D_{EC}) + (1 - w) p_0(\theta_0),
$
which mixes a MAP prior with a noninformative prior. Here, $w$ represents the prior probability that the RCT and EC data are compatible or exchangeable. Mixture priors are appealing because they are conceptually intuitive and highly flexible, allowing the informative component to be tailored. For example, the informative component may be a fully informative prior or a discounted prior, such as a power prior, if one wishes to cap the maximum amount of information borrowed from the EC data. In addition, mixture priors often yield closed-form posteriors, making them computationally simple to implement. However, specifying the mixing weight $w$ has been challenging in practice. Importantly, robustness should not be confused with adaptive information borrowing. Because the full information contained in the mixture prior is incorporated into the posterior by Bayes’ rule, the amount of information borrowed via the mixture prior is not truly adaptive. In other words, once $w$ is specified, the amount of prior information is fixed and will be fully incorporated into the posterior.
What the mixture prior such as rMAP prior provides is a heavy-tailed prior distribution that is more robust to prior–data conflict, not dynamic borrowing.

To enable dynamic borrowing according to the compatibility between the RCT and EC data, \citet{yang2023sam} adopted an empirical Bayes framework and developed the \textbf{self-adapting mixture (SAM) prior}. It determines the mixing weight $w$ using likelihood ratio statistics or Bayes factors in a data-driven manner. The SAM prior enables dynamic information borrowing, achieves desirable finite- and large-sample properties including information-borrowing consistency, and remains computationally simple and calibration-free.

Even when outcome-based borrowing is used, exchangeability between the RCT and EC populations may not hold across all individuals. Therefore, recent methods aim to identify subsets of patients that are exchangeable with the RCT population and borrow information selectively. \citet{alt2024leap,alt2025control} developed the Latent Exchangeability Prior (LEAP), in which observations from the RCT and EC data are classified as exchangeable or non-exchangeable through a mixture model. 
\cite{scott2026borrowing} proposed a semiparametric Bayesian time-to-event model that achieves robust dynamic borrowing via lump-and-smear commensurate priors and a flexible ensemble-based baseline hazard. \citet{ohigashi2025nonparametric} proposed identifying clusters of EC patients that are homogeneous with the RCT population using a dependent Dirichlet process mixture model. Similarly, \citet{bi2023pam} proposed the Plaid Atoms Model, a Bayesian nonparametric approach that detects overlapping and unique subpopulations across EC and RCT patients, and borrows information for the common subgroup through power priors. \citet{pan2024bayesian} developed a Bayesian framework assigning priors over subsets of ECs and adaptively integrating the most compatible ones. 
\citet{zhu2026leveraging} proposed the Discounting Individuals Power Prior (DIPP), enabling subject-level discounting via individual-specific power priors with both local and global discounting parameters.
These approaches embody the same philosophy of selective borrowing discussed in Section~\ref{sec:freq}. The challenges for selective borrowing methods, including mixture models and nonparametric clustering, lie in their limited power to assess exchangeability, especially when RCT data are sparse.

A related issue is covariate shift, where the covariate distributions differ between EC and RCT populations. In such cases, a common strategy is to adopt a \textbf{propensity score (PS)-integrated approach} to align the covariate distribution of the EC data with that of the RCT, and then incorporate the adjusted EC data through an informative prior \citep{wang2019propensity, zhao2025ps}.
See Section~\ref{sec:mat} for more discussion. However, these approaches are not fully Bayesian and do not propagate the uncertainty associated with estimating the propensity scores.
\textbf{Regression-based approaches} address this limitation by incorporating covariates directly into the outcome model, for instance, regression-based power priors \citep{chen2000power} or regression-based commensurate priors \citep{hobbs2012commensurate}. A complication, however, is that once patient-level covariates enter the outcome model, such as in regression-based power priors or commensurate priors, the inferential target shifts to a CATE. In contrast, two-stage propensity score–based approaches typically retain a marginal ATE target because the prior is applied to a matched or reweighted subset of outcomes without modeling covariates explicitly in the outcome regression.

The recent FDA draft guidance on the use of Bayesian methods in clinical trials discusses several settings for information borrowing \citep{FDA2026}. For subgroup borrowing, such as in basket trials, Bayesian hierarchical models may be used to borrow information across related populations (see Section~\ref{sec:Bayesian_shrinkage}). For HCTs, the guidance discusses informative-prior approaches for incorporating EC data, including \textbf{power priors}, \textbf{commensurate priors}, \textbf{mixture priors}, and \textbf{elastic priors}. Specifically, it emphasizes that the construction of the prior, the relevance and quality of the borrowed data, and sensitivity analyses for prior–data conflict should be prespecified and transparently justified. For regulatory decision-making, the guidance distinguishes designs that calibrate Bayesian success criteria to control the frequentist type I error rate from those that rely on direct interpretation of posterior probabilities. Correspondingly, operating characteristics should be evaluated via simulation, including type I error, power, bias, mean squared error, and interval performance for calibrated designs, and Bayesian power and the probability of correct decisions under a range of plausible design priors and outcome-drift scenarios for non-calibrated designs.

\paragraph{Shrinkage approaches} \label{sec:Bayesian_shrinkage} Besides informative-prior approaches, \textbf{Bayesian hierarchical modeling} provides a general framework for borrowing information across studies or subgroups. Depending on the application, these subgroups may correspond to different studies or data sources, disease types (e.g., in basket trials), or control populations. In the context of HCTs, since the EC and RCT control arms receive the same treatment, BHM can be readily applied to borrow information from the EC, especially when the assumption that the EC is exchangeable with the RCT control is plausible. Several commonly used methods are closely related to this framework; for example, the MAP prior, commensurate prior, and some power-prior formulations may be viewed as special cases or alternative representations of hierarchical borrowing. When exchangeability across subgroups is doubtful, more robust extensions have been developed, including cluster-based methods \citep{chu2018blast, hobbs2018bayesian, jiang2021shotgun, jiang2021optimal} that identify relatively homogeneous subgroup sets and calibrated approaches \citep{chu2018bayesian} that adaptively regulate the degree of borrowing according to the extent of heterogeneity. A more detailed discussion is deferred to the Supplementary Material Section~D. A summary of the major Bayesian approaches for EC borrowing, including their prior formulations and key borrowing mechanisms, is provided in Supplementary Material Section~E, Table~S1.

\vspace{-3pt}

\subsubsection{Borrowing under outcome drift: Frequentist approach}
\label{sec:freq}

\paragraph{Prognostic adjustment} \label{sec:prog}
Prognostic adjustment provides a simple way to incorporate ECs while requiring minimal assumptions for consistency \citep{schuler2022increasing,gagnon2023precise,vanderbeek2023bayesian,liao2025prognostic,hojbjerre2025powering}. The initial idea is to use the EC data to train a prognostic score model and construct a ``super-covariate'' $W = \check{\mu}_0(X)$. This variable is then included in the RCT-only estimator for $\theta_0$, $\hat\theta_0^{\text{PA}} =\ntrial^{-1}\sum_{i=1}^n S_i\left[\tilde{\mu}_{10}(X_i,W_i)+\frac{1-A_i}{1-e(X_i)}\{Y_i-\tilde{\mu}_{10}(X_i,W_i)\}\right]$, where $\tilde{\mu}_{10}$ is estimated using only RCT data. Consistency of $\hat\theta_0^{\text{PA}}$ does not require exchangeability of EC outcomes, and type I error control and coverage are preserved even if the prognostic model is misspecified \citep{hojbjerre2025powering}.
If ECs differ substantially from the RCT population, the prognostic score may even be unrelated to the outcome in the RCT sample, and including it may increase finite-sample variance, motivating variable selection \citep{liu2025coadvise}.
Recent work uses pooled data to estimate the nuisance outcome model $\mu_{10}(x)$, which is then plugged into the AIPW estimator \eqref{eq:rctom} and optimally combined with the RCT-only estimator, yielding procedure that is at least as efficient as the RCT benchmark \citep{karlsson2024robust}. 
\citet{ma2026integrating} propose a unified calibration framework for integrating external information under covariate-adaptive randomization through an information proxy vector.
More broadly, the prognostic score can be constructed using any predictive model, including deep learning or foundation models \citep{de2025efficient}, providing flexibility while avoiding exchangeability assumptions on ECs.
Although simple and posing no assumptions on ECs, the limitations are that (i) the efficiency gain in finite samples may be small since $\hat\theta_0^{\text{PA}}$ only averages over the RCT sample and EC data are used only for nuisance estimation, and (ii) asymptotically it achieves the same efficiency bound as the RCT-only AIPW. 

\vspace{-2pt}

\paragraph{Test-then-pool}
Test-then-pool is a common strategy for integrating ECs based on pretesting \citep{viele2014use,yang2023elastic,dang2022cross}. The idea is to first assess whether ECs are comparable to RCT controls by testing $H_0^{\rm EC}: \mathbb{E}[Y(0)\mid S=1,X]=\mathbb{E}[Y(0)\mid S=0,X]$. If $H_0^{\rm EC}$ is not rejected, EC and RCT controls are pooled; otherwise, the analysis reverts to the RCT alone. This approach is intuitive and can yield efficiency gains when exchangeability holds, but it has several limitations. The pretest introduces post-selection randomness, so valid inference must account for the testing step to control type I error \citep{yang2023elastic,dang2022cross,chiam2025selection}. With small RCT samples, the test may lack power to detect EC bias, leading to inappropriate pooling, while rejection discards EC data entirely, reducing efficiency. The binary decision to pool or not also ignores gradations of outcome drift and cannot optimally balance bias and variance. Recent work proposes refinements such as equivalence testing \citep{yuan2019design,li2020revisit,xu2025two,yang2026data}, adaptive lasso \citep{li2023frequentist}, adaptive acceptance thresholds \citep{yang2023elastic}, and ESCV-TMLE, which selects between pooled and RCT-only TMLE based on estimated mean squared error and averages the selection across sample splits \citep{dang2022cross}.

\paragraph{Selective borrowing}\label{sec:selbor}
Instead of an all-or-nothing decision, selective borrowing adaptively incorporates only EC units that satisfy Assumption~\ref{ass:y}. Like matching, the assessment is made at the unit level, allowing biased ECs to be excluded while retaining comparable ones; however, matching primarily addresses covariate shift, whereas selective borrowing addresses outcome drift. For an EC unit $j$ with $S_j=0$, the bias relative to the RCT controls can be expressed as $b_j = Y_j - \mathbb{E}[Y(0)\mid S=1,X_j] = Y_j - \mathbb{E}[Y\mid S=1,A=0,X_j]$, which is identifiable using the RCT control arm. Selective borrowing methods estimate or test these unit-level biases and retain only EC units with negligible bias. \citet{gao2025improving} proposed the penalized selective borrowing augmented calibration weighting (PSB-ACW) estimator, which estimates biases and applies adaptive lasso penalties to shrink negligible biases toward zero, retaining conditionally exchangeable ECs while discarding incompatible ones. \citet{gao2025doubly} extended PSB-ACW to time-to-event outcomes using DR-learners for initial bias estimation. PSB-ACW enjoys double robustness, local efficiency, and asymptotic selection consistency but relies on moderate to large samples. \citet{zhu2025enhancing} proposed conformal selective borrowing (CSB), which uses distribution-free conformal $p$-values to test conditional exchangeability of individual EC units and constructs a CSB-AIPW estimator with targeted tuning to balance bias and variance; combined with Fisher randomization tests (FRT), this approach provides exact finite-sample type I error control under the sharp null (see Section~\ref{sec:ri}) and can improve power when EC bias is negligible or detectable. Extensions to binary outcomes have been developed with general effect measurement and tailored conformal scores \citep{liu2025robust}. \citet{yang2025adaptive} proposed an influence-function method that selects EC units according to their perturbation of the outcome model. Overall, selective borrowing approaches borrow partial information and provide safeguards against bias, but also highlight the difficulty of identifying exchangeable subsets when the RCT control sample size is limited, where post-selection valid inference is needed.

\paragraph{Bias-model-based integration}
\label{sec:bm}
If all ECs have non-zero bias, instead of selecting none of them, bias-model-based methods correct their bias relative to RCT controls. Borrowing is achieved by estimating and removing this bias so that adjusted EC outcomes can contribute information without compromising validity. One strategy augments fixed- or random-effects models with a parameter capturing the discrepancy between RCT and EC controls \citep{kotecha2024leveraging}. This summarizes bias as a single shift or variance component and performs well when the RCT sample is small, but it can be restrictive when bias varies with covariates or the model is misspecified. A more flexible strategy models the bias function $\delta(x)=\mathbb{E}[Y(0)\mid S=0,X=x]-\mathbb{E}[Y(0)\mid S=1,X=x]=\mathbb{E}[Y\mid S=0,X=x]-\mathbb{E}[Y\mid S=1,A=0,X=x]$, which captures conditional differences in control outcomes between EC and RCT populations and is identifiable from observed outcomes. The function $\delta(\cdot)$ can be estimated using parametric regression \citep{stuart2008matching,kallus2018removing,li2023confounding,yang2024datafusion}, partial knowledge or restrictions on the bias function \citep{wu2025comparative,li2024efficient,parikh2025cautionary}, penalized methods for high-dimensional covariates \citep{cheng2023enhancing,gu2024incorporating,ye2025integrative,zhang2025addressing}, sieve or other nonparametric estimators \citep{mao2025statistical}, or modern machine learning and adaptive targeted learning approaches \citep{wu2022integrative,van2024adaptive}. \cite{zhou2021incorporating} incorporated Bayesian Additive Regression Trees to flexibly estimate the bias while regularizing the bias to encourage borrowing in finite samples. A key challenge is that when the bias function is complex and must be estimated flexibly, most information is spent learning $\delta(\cdot)$ rather than the estimand, so ECs may add little efficiency; nevertheless, retaining the option of no efficiency gain is important for preserving type I error in such scenarios.

\paragraph{Estimator averaging}  
Estimator averaging approaches combine RCT-only and EC-augmented estimators with data-adaptive weights, formalizing the bias–variance tradeoff: compatible EC-augmented estimator receives more weight for efficiency, while questionable compatibility shrinks toward the RCT-only benchmark. From a minimax perspective, \citet{chen2021minimax} showed that adaptive combination can achieve near-optimal risk for point estimation but generally cannot deliver shorter valid confidence intervals without additional assumptions. \citet{cheng2021adaptive} minimized estimated mean squared error in a penalized framework that preserves consistency under bias while recovering efficiency when exchangeability holds, and shrinkage estimators based on unbiased risk estimation \citep{rosenman2023combining} provide bounded worst-case error and asymptotic optimality. \citet{yang2025cross} determine borrowing levels through cross-validated predictive performance. \cite{wu2025pessimistic} frame integration as pessimistic policy learning, learning covariate-dependent borrowing weights by minimizing an upper bound on MSE and achieving oracle-adaptive performance across heavy-tailed rewards and distributional shifts. \citet{han2024improving} propose a James–Stein–type shrinkage method combining study-level regression coefficients when only external summary statistics are available. \citet{schwartz2023harmonized} develop harmonized estimators that align EC-augmented subgroup effects with the overall RCT estimate. Risk-diagnostic approaches such as performance curves and bias thresholds \citep{oberst2022understanding} help identify when borrowing improves upon the RCT benchmark. Overall, estimator averaging offers a robust adaptive strategy that automatically down-weights incompatible ECs and achieves near-optimal risk, but because aggregation occurs at the estimator level, it cannot perform unit-level selection of ECs and thus may miss efficiency gains when only a subset of ECs is exchangeable. Moreover, achieving valid inference with efficiency gains for adaptively averaged estimators remains challenging \citep{chen2021minimax}.

\paragraph{Randomization inference}\label{sec:ri}
Randomization inference uses the randomization mechanism itself as the basis for statistical inference \citep{fisher1935}, providing finite-sample exact $p$-values for any test statistic and is widely endorsed \citep{rosenberger2019randomization,carter2024regulatory}. It is particularly useful in small-sample trials and complex designs \citep{simon2011using,plamadeala2012sequential}, and appears in regulatory guidance to ensure type I error control in adaptive settings when large-sample approximations may fail \citep{fda2019ada}. Because HCTs include randomization, this framework applies naturally. To test the \textit{sharp null} in the RCT population, $Y_i(1)=Y_i(0)$ for $i\in\{i:S_i=1\}$, the Fisher randomization test (FRT) compares the observed statistic $T(\boldsymbol{A})$ with its randomization distribution $T(\boldsymbol{A}^*)$, where $\boldsymbol{A}^*$ follows the same assignment distribution as $\boldsymbol{A}=(A_1,\ldots,A_n)^{\text T}$ and permutations are conducted only within the RCT, keeping EC assignments fixed according to the ``analyze as you randomize'' principle \citep{rosenberger2015randomization}. The Fisher $p$-value is $\mathbb{P}_{\boldsymbol{A}^*}(|T(\boldsymbol{A}^*)|\ge |T(\boldsymbol{A})|)$. A key advantage is that type I error is controlled for \textit{any} statistic $T$, allowing EC information to be incorporated through estimators such as CSB-AIPW (Section~\ref{sec:selbor}) \citep{zhu2025enhancing}, Bayesian approaches \citep{ren2025leveraging}, or other borrowing estimators, though power depends on the choice of $T$ and may decline if biased EC data are borrowed \citep{zhu2025enhancing}. A practical concern is computational cost due to repeated evaluation of $T$ across randomizations, although modern computing has largely mitigated this burden \citep{bind2020,efron2021computer}. Inverting FRT yields Hodges--Lehmann estimators and randomization-based confidence intervals; see Section~2.4 of \cite{rosenbaum2020design} and recent developments \citep{luo2021leveraging,zhu2023pair,fiksel2024exact,zhu2024rejoinder,aronow2026randomization}. Overall, randomization inference in HCTs provides finite-sample type I error control and accommodates selection uncertainty in dynamic or selective borrowing, though the classical FRT targets the sharp null; recent extensions address weak null hypotheses using studentized statistics \citep{wu2021randomization} and bounded nulls or treatment effect quantiles \citep{caughey2023}.

\subsection{Sensitivity analyses}  
\label{sec:hct-sen}  

We refer to Section~\ref{sec:SAT-sen} for general sensitivity analyses and focus here on sensitivity to Assumption~\ref{ass:y} in HCTs. This assumption is, in principle, testable in HCTs because the RCT provides an internal control arm, and regulatory evaluations typically examine whether outcomes or trends are similar between the RCT control and the EC. However, in finite samples, such comparisons may be noisy, leaving residual uncertainty.

\begin{table}[t]
\centering
\caption{Software for hybrid RCT and EC methods.}
\label{tab:hybrid_software}
\small
\setlength{\tabcolsep}{4pt}

\resizebox{\textwidth}{!}{%
\begin{tabular}{
p{5.0cm}
p{2.8cm}
p{3.0cm}
p{2.2cm}
p{3.0cm}
p{3.0cm}
}
\toprule
\textbf{Name [version if available]} &
\textbf{Outcome$^1$} &
\textbf{Documentation} &
\textbf{\# dep$^2$} &
\textbf{First release$^3$} &
\textbf{Last updated} \\
\midrule

\href{https://github.com/shuyang-stat/IntegrativeHTECf}{IntegrativeHTECf [0.1.0]} &
Cont/Bin &
\href{https://github.com/shuyang-stat/IntegrativeHTECf}{README} &
4 &
Jul 17 2020 &
Jul 17 2020 \\

\href{https://github.com/XiLinStats/ManyData}{ManyData [0.0.1.9001]} &
Cont/Bin &
\href{https://github.com/XiLinStats/ManyData}{README} &
7 &
Apr 20 2023 &
Mar 06 2024 \\

\href{https://github.com/ldliao/tl4rct}{tl4rct} &
Cont/Bin &
\href{https://github.com/ldliao/tl4rct}{README} &
N/A &
May 30 2023 &
Nov 08 2023 \\

\href{https://github.com/Gaochenyin/ElasticIntegrative}{ElasticIntegrative [1.1]} &
Cont/Bin &
\href{https://github.com/Gaochenyin/ElasticIntegrative/tree/main/vignettes}{Vignettes} &
7 &
Jun 23 2023 &
Sep 10 2023 \\

\href{https://github.com/tq21/atmle}{atmle [0.1.0]} &
Cont/Bin &
\href{https://github.com/tq21/atmle}{README} &
7 &
Jul 30 2023 &
Jun 15 2025 \\

\href{https://cran.r-project.org/web/packages/hdbayes/index.html}{hdbayes [0.1.1]} &
Cont/Bin &
\href{https://cran.r-project.org/web/packages/hdbayes/vignettes}{Vignettes} &
10 &
Apr 08 2024 &
Aug 22 2024 \\

\href{https://github.com/RickardKarl/IntegratingExternalControls}{IntegratingExternalControls} &
Cont/Bin &
\href{https://github.com/RickardKarl/IntegratingExternalControls}{README} &
N/A &
Jun 24 2024 &
Dec 12 2024 \\

\href{https://github.com/IntegrativeStats/intFRT}{intFRT [0.3]} &
Cont/Bin &
\href{https://github.com/IntegrativeStats/intFRT/tree/main/vignettes}{Vignettes} &
7 &
Oct 09 2024 &
May 22 2025 \\

\href{https://cran.r-project.org/web/packages/postcard/index.html}{postcard [1.0.1]} &
Cont/Bin &
\href{https://cran.r-project.org/web/packages/postcard/vignettes/}{Vignettes} &
18 &
Apr 08 2025 &
Jul 01 2025 \\

\href{https://cran.r-project.org/web/packages/RBesT/index.html}{RBesT [1.8-2]} &
Cont/Bin &
\href{https://cran.r-project.org/web/packages/RBesT/vignettes}{Vignettes} &
19 &
Apr 25 2025 &
Apr 25 2025 \\

\href{https://cran.r-project.org/web/packages/psrwe/index.html}{psrwe* [3.2]} &
Cont/Bin/TTE &
\href{https://cran.r-project.org/web/packages/psrwe/vignettes}{Vignettes} &
10 &
Sep 08 2020 &
Jul 16 2024 \\

\href{https://cran.r-project.org/web/packages/psborrow2/index.html}{psborrow2 [0.0.4.0]} &
Cont/Bin/TTE &
\href{https://cran.r-project.org/web/packages/psborrow2/vignettes/}{Vignettes} &
10 &
Jun 04 2021 &
Feb 12 2025 \\

\href{https://cran.r-project.org/web/packages/EScvtmle/index.html}{EScvtmle [0.0.2]} &
Cont/Bin/TTE &
\href{https://cran.r-project.org/web/packages/EScvtmle/refman/EScvtmle.html}{Ref manual} &
8 &
Nov 02 2022 &
Jan 05 2023 \\

\href{https://github.com/Gaochenyin/SelectiveIntegrative}{SelectiveIntegrative [2.0]} &
Cont/Bin/TTE &
\href{https://github.com/Gaochenyin/SelectiveIntegrative}{README} &
4 &
Jul 04 2023 &
May 14 2025 \\

\href{https://cran.r-project.org/web/packages/SAMprior/index.html}{SAMprior [2.0.0]} &
Cont/Bin/TTE &
\href{https://cran.r-project.org/web/packages/SAMprior/vignettes/}{Vignettes} &
6 &
Aug 15 2023 &
Jan 17 2025 \\

\href{https://cran.r-project.org/web/packages/beastt/index.html}{beastt [0.0.3]} &
Cont/Bin/TTE &
\href{https://cran.r-project.org/web/packages/beastt/vignettes/}{Vignettes} &
17 &
Jun 20 2024 &
May 05 2025 \\

\href{https://cran.r-project.org/web/packages/BayesPPD/index.html}{BayesPPD [1.1.3]} &
Cont/Bin/TTE &
\href{https://cran.r-project.org/web/packages/BayesPPD/vignettes/}{Vignettes} &
1 &
Jan 13 2025 &
Jan 13 2025 \\

\href{https://cran.r-project.org/web/packages/brms/index.html}{brms [2.23.0]} &
Cont/Bin/TTE &
\href{https://cran.r-project.org/web/packages/brms/vignettes/}{Vignettes} &
25 &
Sep 09 2025 &
Sep 09 2025 \\

\href{https://github.com/pathwayrf/rdborrow}{rdborrow [0.0.1.0]} &
Longitudinal &
\href{https://github.com/pathwayrf/rdborrow/tree/main/vignettes}{Vignettes} &
12 &
Apr 04 2024 &
Aug 28 2024 \\

\bottomrule
\end{tabular}
} % end resizebox

\vspace{0.5em}
\raggedright
\footnotesize
{$*$ Indicates relevance to SAT+EC settings. \\
$^1$ Outcome abbreviations: Cont = Continuous, Bin = Binary, TTE = Time-to-Event.\\
$^2$ Number of dependents or imports \\
$^3$ First activity or release, for psrwe based on [1.2] and for psborrow2 based on psborrow [0.1.0]
}
\end{table}

Methods in Section~\ref{sec:ps} nevertheless rely on Assumption~\ref{ass:y} to improve efficiency, and sensitivity analysis offers a complementary framework that evaluates how inferences change under controlled violations of Assumption~\ref{ass:y}. \citet{yi2023testing} proposed a ``test-twice'' procedure that compares treatment effect estimates from RCT-only and RCT+EC analyses and reverts to the RCT benchmark when a significant discrepancy suggests EC bias. A key limitation from a scientific and regulatory perspective is that a non-significant result does not necessarily imply negligible bias.
\citet{gordon2025non} developed a nonparametric sensitivity analysis introducing a parameter that bounds deviations between $\mathbb{E}[Y(0)\mid S=0,X]$ and $\mathbb{E}[Y(0)\mid S=1,X]$, constructing bias-aware confidence intervals valid under any specified bound, consistent with the framework of \citet{diaz2013sensitivity}, where the causal gap is varied and summarized by a G-value representing the magnitude required to overturn conclusions. 
\cite{lin2026introducing} proposed b-value framework to construct confidence intervals indexed by the magnitude of potential bias when combining unbiased and biased estimators and reports the maximum relative bias at which the statistical conclusion would change.
\citet{liu2026value} proposed an E-value–informed sensitivity analysis for HCTs based on two metrics, the HC-value and RD-value, which quantify the confounding strength required to overturn the estimated treatment effect and benchmark it against the residual imbalance between internal and external controls.
\citet{lanners2025data} adopt a partial identification perspective, characterizing identified sets for treatment effects under weaker assumptions that allow residual bias between RCT and external data yielding interval estimates guaranteed to contain the true effect but often wider. 
Overall, sensitivity analysis in HCT quantifies robustness to a range of plausible departures from the assumptions. It highlights the tradeoff between validity and efficiency and provides a principled way to evaluate findings when EC exchangeability is uncertain.

\subsection{Software}

Table~\ref{tab:hybrid_software} lists 19 R packages available on CRAN or GitHub: 10 focus on binary or continuous outcomes, 8 also support time-to-event (TTE) outcomes, and one targets longitudinal outcomes. For packages with available vignettes, links are provided. We also report the last update date, the number of dependencies/imports, and the date of first activity or release. The number of dependencies is reported because each may require validation in regulatory settings and indicates potential evaluation and maintenance complexity.

\section{Discussion}
\label{sec:con}

ECTs occupy an increasingly important role in modern clinical research, particularly in therapeutic areas where evidence from traditional randomized trials is difficult to obtain due to ethical considerations. The methodological literature has expanded rapidly across causal inference, Bayesian dynamic borrowing, and hybrid trial design, which, while valuable, can obscure the conceptual structure underlying different approaches. In this paper, we organize ECT methodology using a scientific roadmap consisting of six components: scientific question, observed data structure, identifiability assumptions, statistical parameters, estimation strategies, and sensitivity analysis. This perspective clarifies that methodological differences arise not only from estimation techniques themselves but also from divergent choices of assumptions, especially regarding exchangeability between trial and external populations. 

Several general lessons emerge. First, randomization remains the gold standard: HCTs benefit from the internal benchmark provided by the RCT for unbiased estimation and for diagnosing discrepancies in ECs, and, especially when confounders are not known a priori, borrowing methods cannot compensate for the absence of randomization when unmeasured confounding is substantial. Second, efficiency and robustness are inherently in tension: methods delivering the largest precision gains typically rely on stronger exchangeability assumptions, whereas approaches that relax these assumptions necessarily sacrifice efficiency through discounting, partial identification, or sensitivity-based inference. Third, sensitivity analysis should be treated as a core inferential component, since key assumptions linking external data to the trial population are untestable in SAT+EC and may be underpowered to test in HCT. Finally, clear communication is essential, statisticians must convey the strengths and limitations of different approaches to industry collaborators and regulators; as designs and tools for borrowing external data proliferate, transparent justification of assumptions, estimands, and design choices remains essential. 

The use of ECs raises several recurring questions \citep{freidlin2023augmenting}, which we discuss as follows.
\textbf{Can the type I error rate be controlled when borrowing from ECs?} The key issue is not whether control is possible but under what assumptions. With clearly defined estimands, appropriate identification conditions, prespecified decision rules, and calibrated procedures, type I error control can be maintained; some approaches, such as prognostic adjustment, require no assumptions beyond those for RCT-only analyses. 
\textbf{If type I error is strictly controlled, are power gains possible?} Although \cite{kopp2020power} suggest this is impossible, their result relies on settings where UMP or UMP-unbiased tests exist, under parametric assumptions. This may not extend to more realistic settings where the causal estimand is defined nonparametrically and estimation is semiparametric with potentially misspecified nuisance functions. In such cases, the methods in Section~\ref{sec:freq} show that power gains are possible.
\textbf{Is dynamic borrowing illusory because the RCT control is too small to detect bias?} Critics have noted that small RCT control arms may lack sufficient power to detect bias in ECs, potentially limiting the effectiveness of dynamic borrowing \citep{galwey2017supplementation, freidlin2023augmenting}, particularly in precision medicine settings targeting small biomarker-defined populations. Indeed, dynamic borrowing procedures, whether Bayesian or frequentist, cannot perfectly detect all forms of incompatibility in small samples. Nevertheless, they can mitigate substantial discrepancies and provide partial protection. Their value should therefore be evaluated relative to realistic alternatives, including no borrowing with limited power or full borrowing without safeguards. In addition, prognostic adjustment methods discussed in Section~\ref{sec:prog} do not require bias detection to achieve consistency, and randomization inference in Section~\ref{sec:ri} provides type I error control even when dynamic or selective borrowing is imperfect. In such settings, careful trial design is essential to assess whether incorporating external data is worthwhile, as choices like unbalanced randomization to facilitate borrowing may reduce efficiency if external information ultimately contributes little through dynamic borrowing.

Several directions merit further investigation. 
(1) \textbf{Complex innovative designs:} An important and underexplored context for ECT methodology is its integration within master protocols and other complex innovative designs \citep{fda2023masterprotocols,FDA_CID_Meeting_Program,FDA2026}. Platform, basket, and umbrella trials evaluate multiple treatments or biomarker-defined subgroups within a shared infrastructure, creating opportunities for borrowing across arms, time, and subpopulations. Adapting ECT methods to these settings, where borrowing occurs across multiple dimensions, remains an important open direction; the scientific roadmap provides a useful framework for clarifying estimands, assumptions, and identification strategies in such designs.
(2) \textbf{Synthetic and AI-generated data:} While related to ECT methodology, extending principled integration to real and synthetic data (e.g., digital twins), as well as tabular foundation models (e.g., TabPFN \citep{hollmann2025accurate,zhang2025tabpfn}), is promising but raises new challenges for validity and efficiency \citep{yang2025integrating}.
(3) \textbf{Bayesian–frequentist synthesis:} Recent methods combine Bayesian and frequentist ideas to leverage the strengths of both paradigms, achieving improved small-sample performance while retaining assumption-lean asymptotic validity. The Bayesian bootstrap has been used for approximate Bayesian inference in dynamic borrowing \citep{wang2025robust}, and calibrated Bayesian framework provides promising direction \citep{rubin1984bayesianly,little2011calibrated,wang2018approximate,breunig2025double}.
(4) \textbf{Trial planning:} borrowing strategies should be incorporated at the design stage rather than treated as post hoc adjustments, with further work needed on sample size calculations and adaptive decision rules under potential borrowing \citep{zhang2022bayesian,bi2023beats,ruan2024electronic,guo2024adaptive,chen2024sequential,tian2025beam,gao2025designing,kojima2026sample,liu2025sample}. 
(5) \textbf{Long-term outcomes with external longer follow-up:} hybrid designs are especially relevant when ECs provide longer follow-up than the randomized trial \citep{jackson2017extrapolating}, such as Phase III studies with treatment crossover after the primary endpoint; combining randomized short-term evidence with external long-term outcomes raises challenges including time-dependent confounding, informative censoring, and calendar-time effects, and recent work has explored difference-in-differences and synthetic control approaches \citep{zhou2024estimating}. 
(6) \textbf{Benchmarking:} despite rapid methodological development, systematic benchmark studies in ECTs remain limited, and reproducible evidence from multiple settings is essential to refine practical guidance \citep{wang2023emulation,lin2024data}.  
(7) \textbf{Data access constraints:} privacy regulations often restrict access to individual-level data, leaving investigators with summary statistics or locally trained models, requiring new methods for valid causal inference under limited data granularity or using federated learning. 
(8) \textbf{Rare disease settings:} rare disease trials frequently rely on external data because randomized studies are unfeasible, but small samples, heterogeneous disease courses, and evolving outcomes amplify bias risks and underscore the need for transparent assumptions and sensitivity analyses. 
(9) \textbf{Regulatory translation:} Methodological advances should be aligned with decision frameworks acceptable to regulators, particularly in quantifying uncertainty under unverifiable assumptions and clearly communicating sensitivity analyses. From a regulatory perspective, key considerations include (i) the use of fit-for-purpose data, meaning that ECs are expected to be sufficiently compatible with trial data; methods that rely on accommodating incompatibility through adaptive borrowing may therefore pose challenges for regulatory adoption; and (ii) pre-specification of methods and analyses to limit risks such as p-hacking and to ensure reproducibility, which in turn requires a clear understanding of operating characteristics. In addition, methods should be interpretable, with assumptions and findings accessible to nontechnical stakeholders, including clinicians, prescribers, and patients; overly complex approaches may hinder regulatory adoption.

\vspace{-5pt}

\section*{Data availability statement}
Data sharing is not applicable to this article as no new data were created or analyzed.

\vspace{-5pt}

\if1\anon
{
\section*{Acknowledgments}
This project is supported by the Food and Drug Administration (FDA) of the U.S. Department of Health and Human Services (HHS) as part of a financial assistance award U01FD007934 totaling \$2,556,429 over three years (subject to the availability of funds and satisfactory progress of the project) funded by FDA/HHS. The contents are those of the authors and do not necessarily represent the official views of, nor an endorsement by, FDA/HHS or the U.S. Government. The authors thank Chenguang Wang for his thoughtful discussion and valuable comments. We used ChatGPT 5 to review grammar and improve the writing.
} \fi

\begin{landscape}
\begin{tiny}

\begin{ThreePartTable}
\begin{TableNotes}
\item $^1$ Efficiency gain over the best RCT-only estimator, which is typically covariate-adjusted, but unadjusted in very small samples with weak covariate prediction.
\item $^2$ Assumption~\ref{ass:y} allows covariate shift while assuming no outcome drift.
\item $^3$ L=Low, M=Moderate, H=High.
\item \textbf{Common estimand comparison}: All methods are evaluated under a common target, the ATE in the RCT population.
\item \textbf{Learner-agnostic comparison}: Comparisons are agnostic to the choice of base learners for outcome or propensity modeling; learner choice is treated as a separate implementation layer as \cite{kunzel2019metalearners}.
\end{TableNotes}

\begin{longtable}{@{}
p{3.6cm}
p{2.5cm}
p{3.1cm}
p{3cm}
p{2cm}
p{1.1cm}
p{3cm}
p{2.8cm}
@{}}
\caption{Summary for the statistical methods (Step 5 in Figure~\ref{fig:ECT}) in hybrid controlled trials.}
\label{tab:hct-sum}\\
\toprule
\textbf{Method class}                  & \textbf{Extra assumptions}                                  & \textbf{Covariate shift handling} & \textbf{Outcome drift handling} & \textbf{Efficiency gain$^1$} & \textbf{Bias risk} & \textbf{Typical use}                                                                                       & \textbf{Software}                             \\* \midrule
\endfirsthead
\multicolumn{8}{c}%
{{\bfseries Table \thetable\ continued from previous page}} \\
\toprule
\textbf{Method class}                  & \textbf{Extra assumptions}                                  & \textbf{Covariate shift handling} & \textbf{Outcome drift handling} & \textbf{Efficiency gain$^1$} & \textbf{Bias risk} & \textbf{Typical use}                                                                                       & \textbf{Software}                             \\* \midrule
\endhead
\bottomrule
\endfoot
\insertTableNotes
\endlastfoot
RCT-only difference-in-means           & --                                                                     & --                               & --                             & --                         & --                 & Benchmark analysis                                                                                               & --                                            \\
RCT-only covariate adjustment          & --                                                                     & --                               & --                             & --                         & --                 & Benchmark analysis                                                                                               & RobinCar, Coadvise                            \\
Sampling propensity score matching & (A\ref{ass:y})$^2$; exact matching                                                  & \yes                             & \no                            & M--H$^3$            & M--H     & Preprocessing; improving balance before other methods; no outcome drift                                          & psrwe                                         \\
Sampling propensity score weighting               & (A\ref{ass:y}); correct sampling model                                           & \yes                             & \no                            & M--H             & M--H     & Moderate EC size; good overlap; no outcome drift                                                                 & intFRT                                        \\
Outcome modeling                     & (A\ref{ass:y}); correct outcome model                                            & \yes (averaging over RCT sample)             & \no                            & H                       & H               & Large EC datasets; smooth outcomes; no outcome drift                                                             & intFRT                                        \\
Doubly robust methods                  & (A\ref{ass:y}); correct sampling or outcome model                                & \yes                             & \no                            & H                       & M           & Primary frequentist approach under no outcome drift                                                              & SelectiveIntegrative, \mbox{intFRT}, rdborrow        \\

Bayesian power priors&
EC and trial control share the same treatment effect parameter& 
\yes (If PS-integrated)&
\yes (Likelihood discount via power parameter)& 
M--H& 
L--M& 
EC data with minimal drift; borrowing can be calibrated to preserve validity& 
BayesPPD\\

Bayesian commensurate priors&
Distribution assumption between EC and trial control treatment effects& 
\yes (If PS-integrated)&
\yes (Shrinkage via commensurate parameter)& 
M--H& 
M& 
EC data with limited outcome drift& 
psborrow2\\

Bayesian elastic priors&
EC and trial control share the same treatment effect parameter& 
\yes (If PS-integrated)&
\yes (Inflating prior variance based on prior-data conflict)& 
M--H& 
L& 
EC data with/without outcome drift; borrowing can be calibrated to preserve validity& 
--\\

Bayesian rMAP/SAM priors&
EC and trial control share the same treatment effect parameter& 
\yes (If PS-integrated)&
\yes (Robustifying borrowing using heavier tail prior (rMAP) or adaptive borrowing via data-driven mixture weight based on prior-data conflict)& 
M--H& 
L& 
Multi-source EC data with limited outcome drift; SAM is able to handle larger outcome drift& 
RBesT, SAMprior\\

Bayesian hierarchical model&
Exchangeability across subgroups or studies& 
\yes (If PS-integrated)&
\yes (Shrinkage based on between-subgroup variation)& 
M--H& 
M& 
Multi-source EC data with exchangeable subgroups& 
brms\\

Clustered-Bayesian hierarchical model&
Exchangeability across subgroups or studies within a latent cluster& 
\yes (If PS-integrated)&
\yes (Latent cluster-driven borrowing among similar subgroups, and shrinkage based on between-subgroup variation)& 
H& 
L& 
Multi-source EC data with exchangeable/non-exchangeable subgroups& 
\url{www.trialdesign.org}\\

Prognostic adjustment                  & --                                                                     & \yes                             & \yes(EC-trained prognostic model for improving nuisance function estimation)                           & L--M              & None               & Safe finite-sample efficiency improvement beyond RCT-only covariate adjustment under unknown EC bias & tl4rct, \mbox{postcard}, \mbox{IntegratingExternalControls} \\
Test-then-pool                         & Selection consistency or valid post-selection inference                & \yes                             & \yes(Test (A\ref{ass:y}); borrow all ECs if satisfied, otherwise none)                           & None or H               & L                & Overall-compatible EC set with protection against obvious outcome drift                                          & ElasticIntegrative, \mbox{EScvtmle}                 \\
Selective borrowing                    & Selection consistency or valid post-selection inference                & \yes                             & \yes(Selectively borrow subset of ECs satisfying (A\ref{ass:y}))                         & M-H                   & L                & A subset of ECs has no outcome drift                                                                             & \mbox{SelectiveIntegrative}, \mbox{intFRT}                  \\
Bias-model-based integration           & Correct bias model                                                     & \yes                             & \yes (Model bias from (A\ref{ass:y}) violation)        & L--M              & L                & ECs have simple-structure outcome drift                                                                          & IntegrativeHTECf, atmle                       \\
Estimator averaging                    & Depends on combination weights                                         & \yes                             & \yes(Averaging unbiased and potentially biased estimators)                           & M                   & M           & Combining unbiased and biased estimators for overall efficiency with moderate bias tolerance                     & --                                            \\

Randomization inference                    & Testing sharp null                                         & \yes                             & \yes(Permute within the RCT and keep EC assignments fixed)                       & L-M                   & None           &     Pursue finite-sample exact type I error rate control for complex test statistics         & intFRT                                            \\

Sensitivity analysis      & Correct sensitivity model                                              & \yes                             & \yes(Borrow all ECs and assess sensitivity to (A\ref{ass:y}) violation)                          & H                       & M           & Overall-compatible EC set with sensitivity analysis for regulatory reporting and high-stakes decisions           & --                                            \\* \bottomrule

\end{longtable}
\end{ThreePartTable}
\end{tiny}
\end{landscape}

\bibliography{ref}

\global\long\def\theequation{S\arabic{equation}}
\setcounter{equation}{0}
\global\long\def\thefigure{S\arabic{figure}}
\setcounter{figure}{0}
\global\long\def\thesection{S\arabic{section}}
\setcounter{section}{0}
\global\long\def\thetheorem{S\arabic{theorem}}
\setcounter{theorem}{0}
\global\long\def\thecondition{S\arabic{condition}}
\global\long\def\theremark{S\arabic{remark}}
\setcounter{remark}{0}
\global\long\def\theproposition{S\arabic{proposition}}
\setcounter{proposition}{0}
\global\long\def\thestep{S\arabic{step}}
\global\long\def\theassumption{S\arabic{assumption}}
\setcounter{assumption}{0}
\global\long\def\thetable{S\arabic{table}}
\setcounter{table}{0}

%\spacingset{1.8} % DON'T change the spacing!

%\vspace{-50pt}

\newpage

\begin{center}

{\large\bf SUPPLEMENTARY MATERIAL}

\end{center}

Section~\ref{sec:bias} summarizes sources of bias in ECs.

Section~\ref{sec:att} reviews statistical design and estimation procedures for ATT in SAT+EC settings. 

Section~\ref{sec:str} discusses additional data structures encountered in data integration. 

Section~\ref{sec:bayes-sh} reviews additional Bayesian shrinkage approaches.

Section~\ref{sec:bayes-table} compares Bayesian approaches for borrowing external control data in hybrid controlled trials, focusing on their prior formulations and borrowing behavior.

\appendix

\section{Sources of bias in ECs}
\label{sec:bias}

Five primary sources of bias in ECs are identified by \cite{fda2019rare}:
\begin{itemize}
    \item \textbf{Selection bias}: Patients from external data to the trial may differ systematically from internal participants in baseline characteristics that influence outcomes.
    \item \textbf{Unmeasured confounding}: Key covariates available in RCTs may be missing in ECs, limiting the ability to adjust for confounding.
    
    \item \textbf{Lack of concurrency}: ECs and RCTs may be conducted in different time periods or care settings, introducing bias due to evolving clinical practices or population characteristics.
    
    \item \textbf{Data quality}: External data may suffer from missing not at random, inconsistent collection procedures, and measurement errors.
    
    \item \textbf{Difference in outcome assessment}: Outcomes may be defined or assessed differently across data sources.
\end{itemize}

In addition, other sources of bias when using ECs include lack of blinding, analytical bias due to insufficient prespecification that may lead to selective reporting, multiple comparisons, or data-driven model selection, and immortal time bias arising from misalignment of time zero between the trial and EC cohorts. These biases must be addressed through careful study design and prespecified analytical strategies to ensure valid and interpretable findings when incorporating ECs into clinical research.

\section{Statistical models, estimation, and inference for ATT in SAT+EC}
\label{sec:att}

We define two key nuisance functions: the \textit{propensity score} $\pi(x)=\mathbb{E}[A\mid X=x]$ \citep{rosenbaum1983central} and the conditional outcome mean $\mu_a(x)=\mathbb{E}[Y\mid A=a,X=x]$ \citep{robins1986new}, also referred to as the \textit{prognostic score} \citep{hansen2008prognostic}.

\subsection{Statistical design} 

\textbf{Covariate selection.} Baseline covariates $X$ can be categorized as (i) \textit{confounders}, associated with both the outcome and sampling; (ii) \textit{precision variables} (or prognostic variables), associated only with the outcome; (iii) \textit{instrumental variables}, associated only with sampling; and (iv) \textit{spurious variables}, unrelated to either. Covariate selection principles follow this taxonomy: confounders should always be included to mitigate bias; precision variables are generally recommended for inclusion to improve efficiency; instrumental variables may reduce efficiency but are sometimes retained to minimize the risk of excluding true confounders. Data-driven approaches have been proposed to balance these considerations \citep{schnitzer2016variable,shortreed2017outcome,yang2020doubly,cho2024variable,colnet2025re}.

\textbf{Overlap assessment and balancing.} Adequate overlap in baseline covariates $X$ between the SAT and EC populations is essential for valid inference. The propensity score $\pi(X)$ is a primary dimension reduction tool; as a balancing score \citep{rosenbaum1984reducing}, it ensures that, conditional on $\pi(X)$, baseline covariates are comparable across populations. In practice, $\pi(X)$ is typically estimated using logistic regression, aiming to achieve covariate balance rather than predictive accuracy. The prognostic score provides another balancing score that summarizes outcome-relevant variation \citep{hansen2008prognostic,yang2023multiply}. Diagnostics include standardized mean differences before and after adjustment, graphical comparisons of estimated propensity score distributions, and assessment of common support \citep{greifer2020covariate}. Limited overlap may indicate practical violations of the positivity assumption~1(iii) and may prompt restriction to regions of common support, reconsideration of the target population \citep{li2018balancing}, or trimming \citep{yang2018asymptotic} in sensitivity analyses.

\subsection{Treatment effect estimation and inference} 

Matching aims to construct an EC sample whose covariate distribution resembles that of treated SAT patients. 
\textit{Nearest-neighbor propensity score matching}, often implemented with a caliper of $0.1$–$0.2$ times the standard deviation of the logit of the propensity score \citep{Austin2011}, is widely used in practice. In hybrid trials or post-market settings with sufficient sample sizes, a tighter caliper (e.g., $0.1$) is typically recommended.
Matching may also incorporate multivariate distance metrics, such as the Mahalanobis distance. We refer to \cite{stuart2010matching} for a comprehensive review. Because the ATT anchors inference on the treated population, matching procedures should preserve the representativeness of the SAT sample and avoid excessive exclusion of treated units.

Under Assumption~1, several estimators arise; we summarize them below and refer to \cite{loiseau2022external,campbell2025doubly} for performance comparisons in SAT settings with ECs. A \textit{matching estimator} compares the average outcome among treated units with that of matched ECs, $\hat{\tau}_{\text{Match}} = n_1^{-1}\sum_{i=1}^n A_i Y_i - n_1^{-1}\sum_{i=1}^n \bone(i\in\mathcal{M}_0)Y_i$, where $\mathcal{M}_0$ denotes ECs matched to treated units. \textit{Inverse propensity weighting} (IPW) estimates the ATT by reweighting EC outcomes, $\hat{\tau}_{\text{IPW}} = n_1^{-1}\sum_{i=1}^n A_i Y_i - n_1^{-1}\sum_{i=1}^n \hat{\pi}(X_i)\{1-\hat{\pi}(X_i)\}^{-1}(1-A_i)Y_i$, where $\hat{\pi}(x)$ estimates ${\pi}(x)$ and consistency requires correct specification of $\hat{\pi}(x)$. \textit{Outcome modeling} (OM) replaces EC outcomes with predicted control outcomes for treated units, $\hat{\tau}_{\text{OM}} = n_1^{-1}\sum_{i=1}^n A_i Y_i - n_1^{-1}\sum_{i=1}^n A_i \hat{\mu}_0(X_i)$, where $\hat{\mu}_0(x)$ estimates $\mu_0(x)$ and consistency requires correct specification of $\hat{\mu}_0(x)$. \textit{Augmented inverse probability weighting} (AIPW) combines both models, $\hat{\tau}_{\text{AIPW}} = n_1^{-1}\sum_{i=1}^n A_i Y_i - n_1^{-1}\sum_{i=1}^n [A_i\hat{\mu}_0(X_i) + \hat{\pi}(X_i)\{1-\hat{\pi}(X_i)\}^{-1}(1-A_i)\{Y_i-\hat{\mu}_0(X_i)\}]$, and is doubly robust, remaining consistent if either $\hat{\pi}(x)$ or $\hat{\mu}_0(x)$ is correctly specified. \textit{Targeted maximum likelihood estimation} (TMLE) refines an initial outcome model through a targeted update using a function of propensity score so that the estimator solves the efficient influence function equation for the ATT; it is doubly robust, respects parameter bounds, and often improves finite-sample performance (see Chapter~8 of \citealp{van2011targeted}). \textit{Double score matching} (DSM) matches on both propensity and prognostic scores, achieving double robustness and improved efficiency \citep{hansen2008prognostic,yang2023multiply,zhang2022practical}. Recent work by \cite{tan2025double} compared several doubly robust ATE estimators and showed that incorporating machine learning, particularly within TMLE, can improve robustness and precision in finite samples.

Both $\pi(X)$ and $\mu_a(X)$ are often estimated using parametric likelihood-based regression models, which remain standard in practice. However, inverse propensity weighting can produce unstable estimates when propensity scores approach zero or one, yielding extreme weights. Covariate-balancing approaches such as calibration weighting address this by directly targeting balance through optimization that minimizes the distance between estimated and uniform weights subject to balance constraints (e.g., \citealp{lee2023improving}). Nevertheless, for ATT, such instability is typically less pronounced, as propensity scores less than one act as a stabilizing factor in ATT weighting. Moreover, unless the treatment group is substantially larger than the control group, which is unlikely in regulatory ECT settings, extreme weights are generally less concerning from a regulatory perspective.

Flexible machine learning methods, including penalized regression, tree-based algorithms, and ensemble learners such as Super Learner \citep{van2007super}, can reduce model misspecification but may introduce overfitting and slower convergence. For doubly robust estimators such as AIPW and TMLE, valid inference with machine learning typically requires cross-fitting or sample splitting to mitigate overfitting bias.

For regular asymptotically linear estimators such as IPW, AIPW, and TMLE, uncertainty could be quantified using influence-function–based estimators.
The nonparametric bootstrap is generally an invalid method for inference when machine learning is used to estimate nuisance parameters \citep{coyle2018targeted}, even when using TMLE or doubly robust machine learning approaches. Instead, one can rely on influence-function–based inference or use the targeted bootstrap.
In small SAT settings, finite-sample performance should be carefully evaluated, as asymptotic approximations may be unreliable.

\section{Alternative scientific questions and data structures in data integration}
\label{sec:str}

\textbf{Transportability, generalizability, or treatment effect heterogeneity assessment.} External data enable inference on questions the RCT was not originally powered to address, requiring careful specification of the target population aligned with the scientific question. For example, inference may target the EC population, leading to generalization problems \citep{degtiar2023review,colnet2024causal,manke2025and,boughdiri2025unified,lee2022doubly,lee2023improving,lee2024transporting,lee2024genrct,yang2022rwd}. Alternatives include the pooled RCT–EC population \citep{li2023improving}, the overlapping subpopulation where their covariate distributions intersect \citep{wang2025integrating,liu2025leveraging}, or averaging the RCT conditional ATE (CATE) over a pooled covariate distribution \citep{van2024adaptive}.
Beyond the ATE, interest may lie in heterogeneous treatment effects \citep{yang2023elastic,yang2024datafusion,wu2022integrative,brantner2023methods,wang2024efficient,d2025modern} and individualized treatment rules \citep{chu2023targeted,wu2023transfer,zhao2025efficient,li2024combining}. Valid inference for the CATE often requires parametric assumptions or flexible machine learning methods such as the Highly Adaptive Lasso \citep{nizam2025highly}.

\textbf{External summary information and federated learning}
In this paper, we focus on scenarios where individual patient-level external data are accessible. In practice, such access may be limited due to privacy or operational constraints. One possibility is that only external summary information (for example, sample moments or estimated effects) can be obtained \citep{chatterjee2016constrained,zhang2020generalized,qin2022selective,hu2022semiparametric,chu2023targeted,zheng2023generalized,
zhai2024integrating,han2024improving,schwartz2024dynamic}. We refer to \cite{chen2024advancing} and \cite{hu2022semiparametric} for a recent review of methods that incorporate summary-level external information while accounting for potential heterogeneity. Another possibility is that patient-level data remain stored locally at each external source site, but each site is able to train a model and share only the resulting fitted parameters. In this setting, federated learning methods can be used to integrate information from external sites to improve estimation at the target site \citep{xiong2023federated,han2025federated,liu2025targeted,cao2025heterogeneity}.

\textbf{External covariate information and semi-supervised learning}
If only covariate information is available externally (that is, the external data are unlabeled), the goal is to improve inference based on the internal labeled data that contain both outcomes and covariates. 
In this case, the observed data can be written as $(S,A,X,SY)$. 
A convenient view is to treat this as a missing-outcome problem under a missing-at-random assumption, allowing standard TMLE methods to be applied. 
This setting is also closely related to semi-supervised learning \citep{zhang2019semi, cheng2021robust, zhang2022high, yu2024data, xu2025unified}.

\textbf{Mismatched covariate set} In practice, two data sources often include different covariates \citep{taylor2023data}. Recent work has addressed these challenges in the context of federated learning \citep{han2023multiply}, transfer learning \citep{chang2024heterogeneous}, representative learning \citep{xu2025representation},
integrating external summary information \citep{han2024improving},
generalization \citep{li2025generalizing,zeng2025efficient} and data fusion for policy learning \citep{williams2025nonparametric}, multi-regional clinical trials \citep{li2026selective}.

\textbf{Auxiliary variables} provide complementary information that can help integrate ECs when direct outcome exchangeability is doubtful.
\textit{Control variates} exploit auxiliary statistics that are correlated with the primary outcome but whose expectation is either known or estimable from the RCT \citep{yang2020combining}. \citet{guo2022multisource} show how such control variates can be constructed by comparing conditional odds ratio estimates across datasets, thereby converting external information into variance-reducing instruments rather than potential sources of bias. \cite{li2024efficient} extends \citet{guo2022multisource}'s strategy to construct semiparametric efficient estimators.
\textit{Negative control outcomes} (NCOs) are outcomes known a priori to be unaffected by treatment but sensitive to the same unmeasured confounding that induces drift between RCT and EC controls. \citet{dang2022cross} embed NCOs into a cross-validated TMLE framework, using discrepancies on NCOs to diagnose hidden bias and adaptively decide whether ECs should be incorporated. This provides a safeguard against incompatibility that cannot be detected by covariates or primary outcomes alone.  
\textit{Secondary outcomes}, routinely measured alongside the primary endpoint, can also stabilize borrowing. \citet{wolf2024leveraging} use multivariate outcome models to leverage correlations between primary and secondary outcomes, showing efficiency gains for subgroup-specific or overall treatment effects. \citet{deng2025new} propose an integrative learning framework that combines multiple secondary outcomes, which is especially useful when primary events are rare or RCT samples are small. The benefit depends on the predictive strength of the secondary outcomes and the validity of the joint outcome model.  

\section{Additional Bayesian shrinkage approaches in HCT}
\label{sec:bayes-sh}

{Bayesian shrinkage approaches}, typically implemented through Bayesian hierarchical modeling (BHM), provide a general framework for borrowing information and are particularly natural when a treatment or intervention is studied across multiple studies, with each study or data source treated as a ``subgroup.’’ Depending on the application, these subgroups may correspond to different studies or data sources, disease types (e.g., in basket trials), or control populations. In the context of HCTs, because the EC and the RCT control arm receive the same treatment, BHM can be naturally applied to borrow information from the EC, particularly when the assumption that the EC is exchangeable with the RCT control arm is plausible.

The core idea of BHM is to jointly model the treatment effects relative to the control, denoted as $\theta_0^{k}$ for $k = 1, \ldots, K$, across $K$ subgroups. Through the hierarchical structure, these subgroup-specific effects can be shrunk toward a common overall effect $\theta_0$, thereby enabling information borrowing across subgroups. This borrowing is justified under the assumption that subgroup effects are exchangeable, that is, similar in distribution and interchangeable before seeing the data. Under this assumption, \citet{thall2003hierarchical} proposed using BHM to adaptively borrow information across subgroups within the same disease, and \citet{berry2013bayesian} considered a similar approach for basket trials. 

However, when the exchangeability assumption is violated, such as in the presence of heterogeneous responses between subgroups (e.g., outcome drift between EC and RCT), BHM can lead to biased estimates, loss of power, or inflated type I error rates \citep{freidlin2013borrowing, chu2018bayesian}. 
These issues are particularly pronounced when the number of subgroups is small (e.g., ten or fewer), where estimation of between-group variability is imprecise and the extent of borrowing can be highly sensitive to hyperparameter choices.
To enhance robustness against violations of the exchangeability assumption, \citet{neuenschwander2016robust} proposed replacing the shrinkage prior with a mixture that combines it with an independent prior for $\theta_0^{k}$ as the second level of the BHM. Similar to the mixture informative prior approach, the main challenge of this method is specifying the mixture weight. \citet{chu2018bayesian} proposed a calibrated BHM that incorporates a calibrated elastic function, similar in spirit to the elastic prior of \citet{jiang2023elastic}, to encourage borrowing among homogeneous subgroups while down-weighting information from heterogeneous ones.

In addition, cluster-based approaches have been explored to partition subgroups that either satisfy or violate the exchangeability assumptions to achieve more accurate information borrowing. For example, \citet{chu2018blast} employ latent class BHM, where subgroups within a latent class are exchangeable, to enhance information borrowing, and leveraged longitudinal biomarkers as auxiliary variables to improve reliability of clustering. 
\citet{hobbs2018bayesian} utilized posterior exchangeability probabilities to evaluate pairwise similarity across subgroups; \citet{jiang2021shotgun} used a Bayesian posterior probability based on response rates and \citet{jiang2021optimal} further extended with utility function that considers the trade-off between type I error and power. One challenge of the cluster-based approach is the limited power for clustering, as the cluster unit is the subgroups, not individual observations. As a result, the improvement can be rather limited.  Using auxiliary variables provides an approach to address this issue \citep{chu2018blast}.

Rather than directly employing a clustering-based approach, a conceptually similar alternative is to utilize Bayesian model averaging to capture various possible configurations of compatibility between EC and the RCT control arm. \citet{kaizer2018bayesian} proposed multisource exchangeability models for this purpose.  \citet{wei2024propensity} further extended this approach by incorporating covariate information through a propensity score weighting strategy. \citet{psioda2021bayesian} adapted the approach for the context of basket trials.

In summary, BHM enables information borrowing across subgroups by shrinking estimates toward a common mean under the exchangeability assumption. While efficient, it can lead to bias or inflated error rates when subgroup heterogeneity is present. Extensions such as calibrated BHMs and clustering-based methods improve robustness by adaptively controlling borrowing but require sufficient within-subgroup data and careful calibration. Related approaches evaluate exchangeability across external sources to mitigate bias in data integration; however, their performance may be sensitive to prior specification, limited sample sizes, and the large space of possible model configurations.

\section{Comparison of Bayesian approaches in HCT}
\label{sec:bayes-table}

Table~\ref{tab:bayes_priors} provides a comparison of key Bayesian approaches for borrowing external control data in hybrid controlled trials, highlighting their prior formulations and corresponding borrowing mechanisms.

{\spacingset{1}\small\renewcommand{\arraystretch}{1.5}
\begin{longtable}{>{\raggedright}p{0.18\textwidth}
                  >{\raggedright}p{0.44\textwidth}
                  >{\raggedright\arraybackslash}p{0.33\textwidth}}
\caption{Bayesian approaches for borrowing external control (EC) data in
hybrid controlled trials. Here $\theta_0$ denotes the RCT control parameter,
$\theta_0^{\mathrm{EC}}$ its EC counterpart, $D_{\mathrm{EC}}$ the EC data, $D_{\mathrm{RC}}$ the RCT control data,
$L(\cdot)$ the likelihood, $p_0(\theta_0)$ a non-informative prior,
$a_0\in[0,1]$ a power/discount parameter, $\eta$ a commensurability parameter,
$w\in(0,1)$ a mixture weight, $T$ a congruence measure, and $g(T)\in(0,1)$ an
elastic function.}
\label{tab:bayes_priors}\\
\toprule
\textbf{Method} & \textbf{Prior formulation} & \textbf{Key borrowing mechanism} \\
\midrule
\multicolumn{3}{l}{\textbf{Informative prior approaches}}\\
\midrule
\endfirsthead
\multicolumn{3}{c}{{\tablename\ \thetable{} -- continued from previous page}} \\
\toprule
\textbf{Method} & \textbf{Prior formulation} & \textbf{Key borrowing mechanism} \\
\midrule
\endhead
\midrule
\multicolumn{3}{r}{{Continued on next page}} \\
\endfoot
\bottomrule
\endlastfoot

\textit{Fully informative prior}
&
$p(\theta_0 \mid D_{\mathrm{EC}}) \propto L(D_{\mathrm{EC}} \mid \theta_0)\, p_0(\theta_0)$.
&
EC information is fully incorporated. When multiple EC datasets are available, the MAP prior can be adopted. \\

\midrule
\textit{Meta-analytic predictive prior} \citep{neuenschwander2010summarizing}
&
Level 1:\; $D_{\mathrm{EC},k} \mid \theta_0^k \sim L(\cdot \mid \theta_0^k)$,\; $k = 1,\ldots,K$
\newline
Level 2:\; $\theta_0^k \mid \mu, \omega^2 \sim \mathcal{N}(\mu,\, \omega^2)$
\newline
Level 3:\; $\mu \sim p_0(\mu)$,\; $\omega^2 \sim p(\omega^2)$
\newline
$p_{\mathrm{MAP}}(\theta_0 \mid D_{\mathrm{EC},1}, \ldots, D_{\mathrm{EC},K})
=
\int
p(\theta_0 \mid \mu, \omega^2)\,
p(\mu, \omega^2 \mid D_{\mathrm{EC},1}, \ldots, D_{\mathrm{EC},K})
\, d\mu\, d\omega^2$
&
The predictive prior serves as a fully informative prior that incorporates information across multiple EC datasets while accounting for between-study heterogeneity. \\

\midrule
\textit{Power prior} \citep{chen2000power}
&
$p(\theta_0 \mid D_{\mathrm{EC}}, a_0) \propto L(D_{\mathrm{EC}} \mid \theta_0)^{a_0}\, p_0(\theta_0)$,\quad $a_0 \in [0,1]$.
&
The EC likelihood is discounted by the power parameter $a_0$; $a_0=1$ achieves full borrowing, whereas $a_0=0$ yields no borrowing. \\

\midrule
\textit{Normalized power prior} \citep{duan2006evaluating}
&
$p(\theta_0 \mid D_{\mathrm{EC}}, a_0) = \dfrac{L(D_{\mathrm{EC}} \mid \theta_0)^{a_0}\, p_0(\theta_0)}{C(a_0)}$,
\newline
$C(a_0) = \int L(D_{\mathrm{EC}} \mid \theta_0)^{a_0}\, p_0(\theta_0)\,d\theta_0$.
&
The normalizing constant $C(a_0)$ ensures a proper posterior; $a_0$ may further be assigned a Beta prior (e.g., $\mathrm{Beta}(1,1)$) to enable data-driven borrowing. \\

\midrule
\textit{Calibrated power prior}  \citep{pan2017calibrated}
&
Same as the NPP, but $a_0 = g(T)$, where $g(T)= 1/ (a+b\exp(T)),$
and $T = T(D_{\mathrm{EC}}, D_{\mathrm{RC}})$ is a pre-specified congruence measure comparing the EC and RCT control data, with $a$ and $b$ calibrated via simulation.
&
When the EC and RCT control data are compatible, $g(T)\to 1$ and full borrowing is recovered; otherwise, $g(T)\to 0$ and no borrowing occurs. \\

\midrule
\textit{Elastic prior} \citep{jiang2023elastic}
&
$p_{\mathrm{E}}(\theta_0 \mid D_{\mathrm{EC}})$ is constructed from $p(\theta_0 \mid D_{\mathrm{EC}})$ with variance inflated by $g(T)^{-1}$:
\newline
$\mathrm{Var}_{\mathrm{E}}(\theta_0) = g(T)^{-1}\,\mathrm{Var}[p(\theta_0 \mid D_{\mathrm{EC}})]$.
&
$g(T)$ is defined similarly to the CPP, based on EC–RCT compatibility.\\
\midrule
\textit{Commensurate prior} \citep{hobbs2011hierarchical}
&
$p(\theta_0 \mid D_{\mathrm{EC}}, \theta_0^{\mathrm{EC}}, \eta) \propto L(D_{\mathrm{EC}} \mid \theta_0^{\mathrm{EC}})\, p(\theta_0 \mid \theta_0^{\mathrm{EC}}, \eta)\, p_0(\theta_0)$,
\newline
$\theta_0 \mid \theta_0^{\mathrm{EC}}, \eta \sim \mathcal{N}(\theta_0^{\mathrm{EC}},\, \eta^{-1})$.
&
Shrinks $\theta_0$ toward $\theta_0^{\mathrm{EC}}$; large $\eta$ induces strong borrowing; small $\eta$ allows less borrowing. \\

\midrule
\textit{Robust MAP (rMAP) prior} \citep{schmidli2014robust}
&
$p(\theta_0 \mid D_{\mathrm{EC}}) = w\, p_{\mathrm{MAP}}(\theta_0 \mid D_{\mathrm{EC}}) + (1-w)\, p_0(\theta_0)$.
&
Borrowing is controlled by $w$; $w=1$ gives full borrowing and $w=0$ gives no borrowing. Default choice of $w = 0.5$ yields a heavy-tailed prior robust to prior–data conflict. \\
\midrule
\textit{Self-adapting mixture prior} \citep{yang2023sam}
&
$p(\theta_0 \mid D_{\mathrm{EC}}) = w(D_{\mathrm{RC}})\, p_{\mathrm{MAP}}(\theta_0 \mid D_{\mathrm{EC}}) + (1-w(D_{\mathrm{RC}}))\, p_0(\theta_0)$,
\newline
$w(D_{\mathrm{RC}}) = \dfrac{R}{1+R} \quad R = \frac{L(D_{\mathrm{RC}} \mid \mathcal{M}_0)}{L(D_{\mathrm{RC}} \mid \mathcal{M}_1)}$,
where $\mathcal{M}_0$ and $\mathcal{M}_1$ indicate the absence and presence of prior--data conflicts.
&
The mixture weight $w(D_{\mathrm{RC}})$ is data-adaptive and calibration-free; when the EC and RCT control data are compatible, $R$ tends to be large and $w\to 1$, yielding full borrowing, whereas under incompatibility, $w\to 0$, yielding little to no borrowing.\\
\hline
\textit{The latent exchangeability prior} \citep{alt2024leap}
&
$f(D_{\mathrm{EC}}, c_0 \mid \theta,\gamma)
=
\prod_{i=1}^{n_0}\prod_{k=1}^K [\gamma_k f(Y_{\mathrm{EC}, i}\mid \theta_0^k)]^{c_{0ik}}$,
with $\theta_0^1 \equiv \theta_0$ denoting the exchangeable component. The induced LEAP is
$p(\theta_0 \mid D_{\mathrm{EC}})
\propto
\int \cdots \int
\sum_{c_0 \in [1,K]^{n_0}}
f(D_{\mathrm{EC}}, c_0 \mid \theta,\gamma)\, p_0(\theta,\gamma)\,
d\theta_0^2 \cdots d\theta_0^K\, d\gamma .$
&
Classifies individual EC units as exchangeable ($c_{0i1}=1$) or not via a latent mixture; only exchangeable units contribute to borrowing.
\\

\midrule
\multicolumn{3}{l}{\textbf{Shrinkage approaches}}\\
\midrule
\textit{Bayesian hierarchical model} \citep{thall2003hierarchical}
&
Level 1:\; $D_k \mid \theta_0^k \sim L(\cdot \mid \theta_0^k)$,\; $k = 1,\ldots,K$
\newline
Level 2:\; $\theta_0^k \mid \theta_0, \omega^2 \sim \mathcal{N}(\theta_0,\, \omega^2)$
\newline
Level 3:\; $\theta_0 \sim p_0(\theta_0),\; \omega^2 \sim p(\omega^2)$.
&
Partial pooling of $K$ source-specific effects toward a common mean; $\omega^2$ governs shrinkage. \\

\midrule
\textit{Calibrated BHM} \citep{chu2018bayesian}
&
As BHM but Level 2 replaced by:
\newline
$\omega^2 = 1 / g(T)$, calibrated via simulation.
&
Down-weights borrowing when subgroup heterogeneity is high, while preserving efficient borrowing when subgroups are homogeneous. \\

\midrule
\textit{Clustered BHM} \citep{jiang2021shotgun}
&
Clusters subgroups by posterior response similarity, then applies separate BHMs within each cluster.
&
Borrowing is restricted to relatively homogeneous clusters, reducing inappropriate borrowing across heterogeneous subgroups. \\

\end{longtable}
}

\end{document}